\documentclass[onecolumn,aps,floats,superscriptaddress,showpacs]{revtex4}
\usepackage{graphicx}
\usepackage{bm}
\usepackage{amsmath}
\usepackage{epsfig}

\newcommand{\beq}{\begin{equation}}
\newcommand{\eeq}{\end{equation}}
\newcommand{\beqa}{\begin{eqnarray}}
\newcommand{\eeqa}{\end{eqnarray}}

\def\gapp{\lower.35em\hbox{$\stackrel{\textstyle>}{\sim}$}}
\def\lapp{\lower.35em\hbox{$\stackrel{\textstyle<}{\sim}$}}
\begin{document}
\bibliographystyle{apsrev}
\title[Review on graphene]{Geometrical and topological aspects of graphene and related materials}
%\article[Review on graphene]{Topical Review}{Geometrical and topological aspects of graphene and related materials}
%\topical[Review on graphene]{Geometrical and topological aspects of graphene and related materials}
\author{A. Cortijo, F. Guinea and M. A. H. Vozmediano}

\address{Instituto de Ciencia de Materiales de Madrid,
CSIC, Cantoblanco, E-28049 Madrid, Spain}
%\ead{vozmediano@icmm.csic.es}
\begin{abstract}
Graphene,
a two-dimensional crystal made of carbon atoms, provides a new and unexpected bridge
between low and high-energy physics. The field has evolved very fast and very good reviews are already available in the literature. 
Graphene constitutes a condensed matter realization of  lower dimensional quantum field theory models that were proposed to confront important -- still unresolved -- puzzles of the area: Chiral symmetry breaking and quark confinement. The new materials named topological insulators, closely related to graphene, are physical realizations of topological field theory.  This article  reviews some of these topics with the aim of bridging the gap and making these condensed matter issues accessible to high energy readers. The electronic interactions in the monolayer are analyzed with special emphasis on the recent experimental confirmation of some theoretical predictions. The issue of spontaneous chiral symmetry breaking in the model materials is also reviewed. Finally we give an extensive description of some recent topological aspects of graphene that allow to understand  the main aspects of topological insulators. 

\end{abstract}

\maketitle
\tableofcontents
\section{Introduction}
The main conceptual advances in modern physics have usually been prompted by almost simultaneous discoveries in different branches. In the past century, statistical physics, quantum field theory (QFT) and condensed matter have had their main developments in parallel with the
best physicists (Feynman, Einstein, Landau, Wigner) contributing
to them all. Cosmology and astrophysics develop
with inputs from particle physics. This fruitful arena has been damaged in the last  half of the past century by the fast development and  specialization of the different fields and by the excess of information and production of scientific papers what has originated a profound gap between different areas.
The experimental realization of graphene,
a two-dimensional crystal made of carbon atoms  in 2004
provides a new and unexpected bridge
between condensed matter and high-energy physics which also involves the fields of elasticity, statistical mechanics, and cosmology.

Graphene is a two-dimensional crystal of carbon atoms arranged in a honeycomb lattice: a single layer of graphite \cite{KN07}. Its synthesis \cite{Netal05,Zetal05}, amazing properties \cite{G09} and potential applications \cite{G09,SMD10}, have  granted the 2010 Nobel prize in Physics  to  A. Geim and K. Novoselov. 
One of the most interesting aspects of the graphene physics from a theoretical point of view is the deep and fruitful relation that it has with quantum electrodynamics (QED) and other quantum field theory  ideas \cite{KN07,S84,H88,GGV94}. The connection arises from the   fact that the low energy excitations of the system can be modeled  by the massless Dirac equation in two spatial dimensions. From the QFT point graphene has given rise to  very interesting developments: the so-called axial anomaly \cite{S84,H88} has acquired special relevance in relation with the recently discussed topological insulators \cite{BHZ06,HK10} which provide a condensed matter realization of the axion electrodynamics \cite{W87,LWetal10}. Charge fractionalization \cite{NS86} has also been explored in the honeycomb lattice with special defects \cite{HCM07,JP07} and quantum field theory in curved space \cite{GGV92,JCV07} and cosmological models  \cite{CV07a,CV07b} have been used to study the electronic properties of the curved material. A very interesting development is associated to the generation of various types of vector fields coming from the elastic properties or from disorder that couple to the electrons in the form of gauge fields \cite{VKG10}. 

Many good reviews are already available in the literature \cite{CGetal07,GSC07,VKG10,AAetal10,KUetal11,CAetal11,SAetal11}, some of them very theoretically oriented \cite{HFA07,Pa09,FW11} so we will focus on some particular aspects that either have not been sufficiently clarified or are the basis of subsequent developments. The topics are also chosen to be in the frontier between condensed matter and high energy physics. 
Under a theoretical point of view the synthesis of graphene has opened a new world where
ideas from different branches of physics can be confronted and tested in the laboratory. On the
electronic point of view it can be shown that the low energy excitations of the neutral system
obey a massless Dirac equation in two dimensions. This special behavior originates on the
topology of the honeycomb lattice and has profound implications to the transport
and optical properties. Although the Fermi velocity is approximately a hundredth of the speed
of light, the masslessness of the quasiparticles brings the physics to the domain of relativistic
quantum mechanics where phenomena like the Klein paradox or the Zitterbewegung can be
explored. None of these questions arise within the quantum field theory approach but its full
applicability to the condensed matter system is questionable. It is also interesting to see how old low dimensional QFT models, particularly the Nambu-Jona Lasinio \cite{NJL61} and the Gross-Neveu \cite{GN74} model, found graphene as a physical realization.

Dirac fermions appeared in condensed matter before the advent of graphene but  are ubiquitous in our days. Again here graphene constitutes the cleanest and one of the most robust examples. 

Topology is a invaluable mathematical tool in many areas of physics, particularly, in high energy and condensed matter physics. In the latter case we can also find apparently disjoint areas where the dynamics of the relevant degrees of freedom in the system are governed by laws based on topological constraints. This is the case for instance, of critical systems in two dimensions comprising the XY spin model, dislocations and melting of solids, liquid crystals and so on \cite{KT73}. Also, the concept of Berry phase will be ubiquitous in the physical properties of crystalline solids, as we will see. In basis of its well known properties, we will use graphene as a laboratory to understand most of the properties derived from the topology associated to Berry phase.
Although the field of topological insulators has followed its own path \cite{HK10}, some of the most fundamental aspects of the the physical phenomena involved can already be found in graphene. Being a very simple material and model, understanding the topological aspects of it opens the way to follow the more sophisticated developments. 

The organization of the article is summarized in the contents. 

\section{A summary of the graphene features}
\label{sec_features}

\subsection{Monolayer graphene.}

The construction of the free action in condensed matter physics proceeds in a very similar
way as in QFT. The non interacting Hamiltonian is determined by the discrete (crystal) and internal symmetries of the system. The ``band theory" provides the dispersion relation of the material and the
electronic properties for a given electron occupancy. In the case of having a metallic system, adding interactions can open a gap and give rise to various non-trivial insulating states \cite{F07} or keep the system gapless within the universal class of the Landau Fermi liquid, the standard model for metals \cite{L57}. (A very important exception is the behavior of one dimensional systems that give rise to a different universality class, the Luttinger liquid \cite{T50,L63}). 

\begin{figure}
\begin{center}
\includegraphics[width=4.5cm]{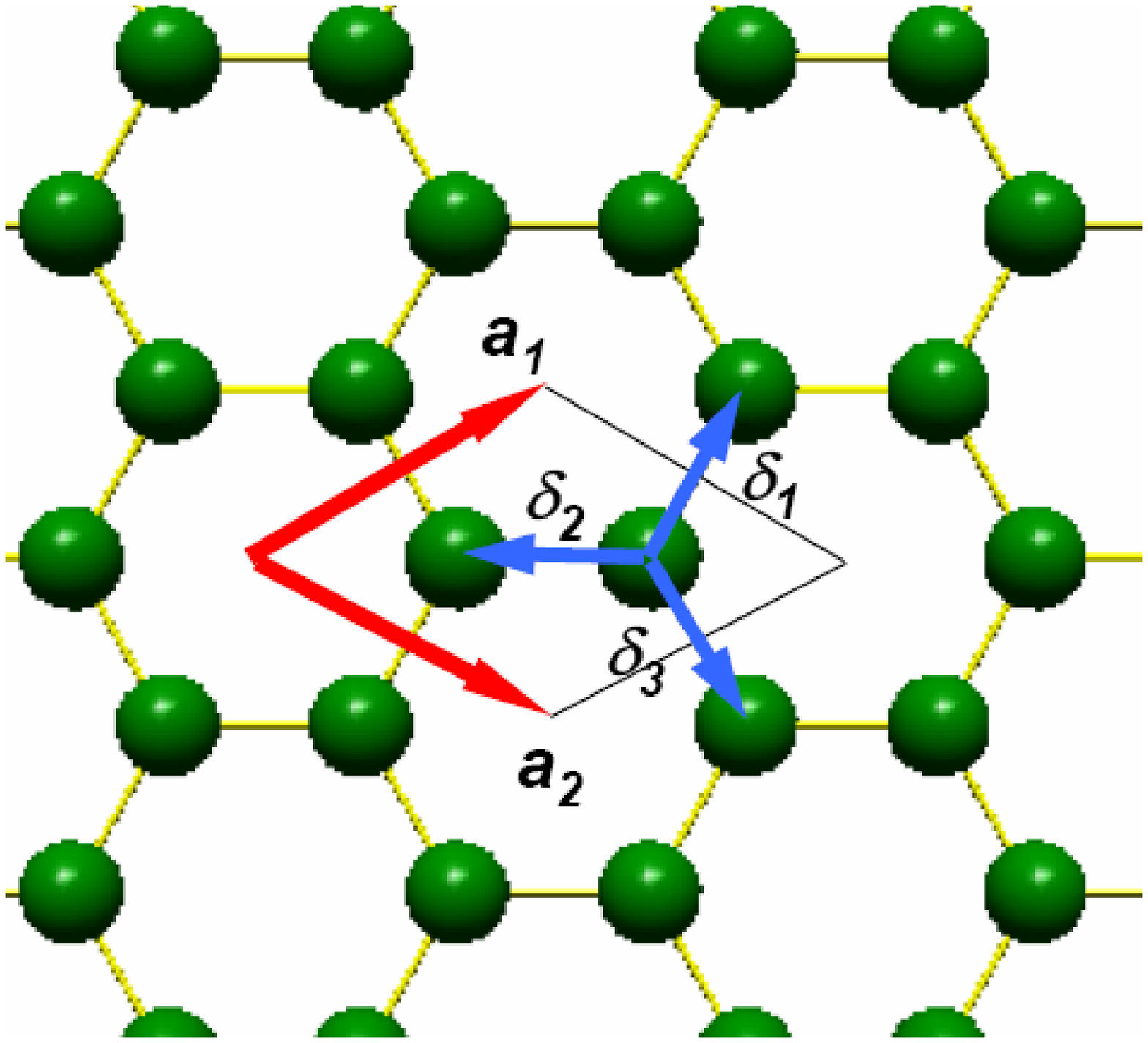}
\hspace{1cm}
\includegraphics[width=6.0cm]{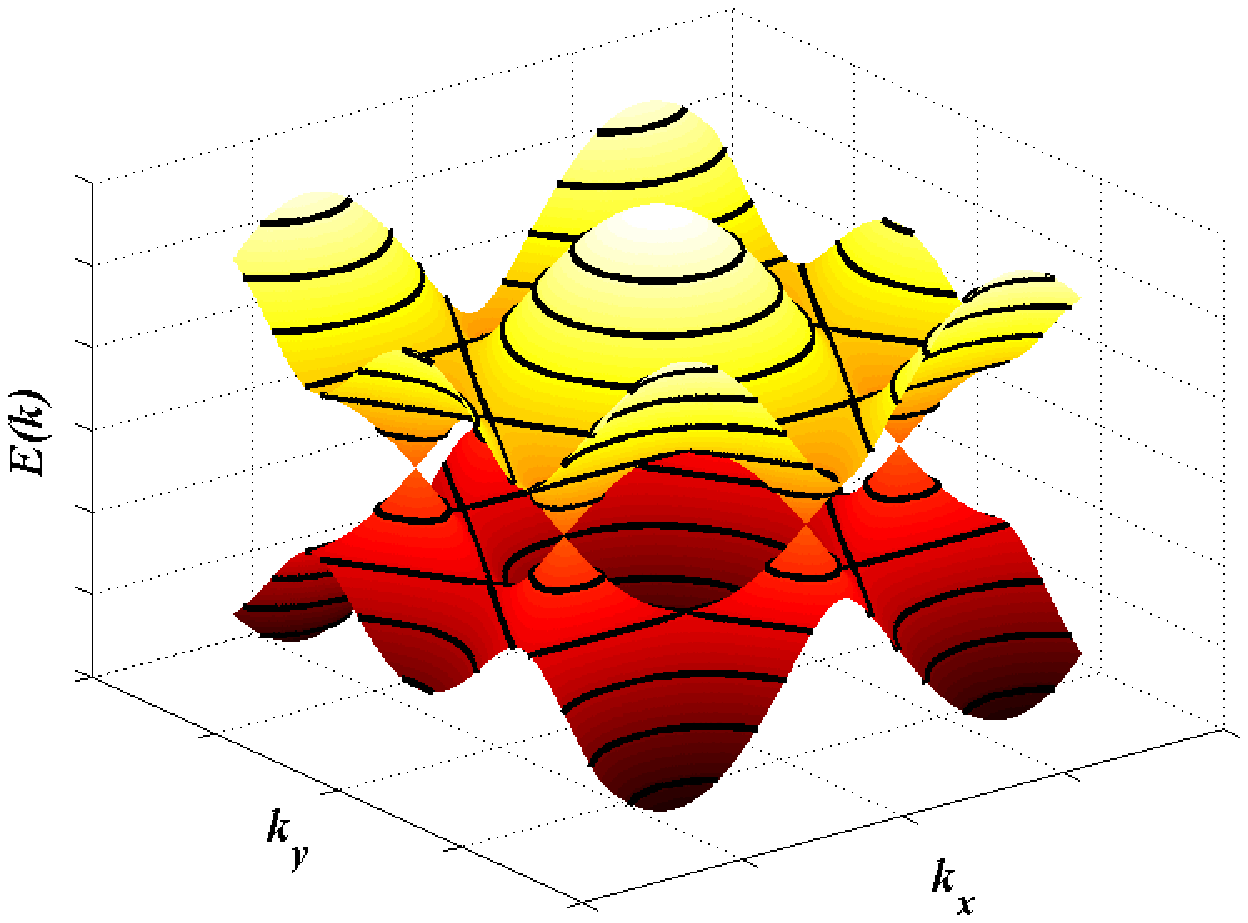}
\caption{Left: The honeycomb lattice is made of two
interpenetrating triangular lattices. Right: the dispersion
relation.}
    \label{lattice}
\end{center}
\end{figure}

The carbon atom has four external $2s^2, 2p^2$ orbitals able to form molecular bonds. 
The crystal structure of graphene consists of a planar honeycomb lattice of carbon atoms shown
at the left hand side of Fig. \ref{lattice}. In the graphene structure the in-plane $\sigma$ bonds are formed from
2s, $2p_x$ and $2p_y$ orbitals hybridized in a $sp^2$ configuration, while the $2p_z$ orbital, perpendicular
to the layer remains decoupled. The $\sigma$ bonds
give rigidity to the structure and the $\pi$ bonds give rise to the valence and conduction bands. The exotic electronic properties of graphene are based on the construction of a model for the $\pi$ electrons sitting in the position of the  Honeycomb lattice drawn by the $\sigma$ bonds. Alternatively, the mechanical properties involve the $\sigma$ bonds with characteristic energies of the order of 7-10 eV. The low energy excitations around the Fermi energy will have  characteristic energies ranging from a few meV up to 1 eV. 

Most of the crystal lattices discussed in text books are Bravais lattices and in two dimensions they can be generated moving an arbitrary lattice point along two defined vector lattices. This happens in the generalized square and triangular lattices in two dimensions.   It is easy to see that this is not the case of the Hexagonal lattice. This lattice is very special: It has the lowest coordination in two dimensions (three) and it has two atoms per unit cell. As it can be seen in Fig. \ref{lattice}, the Hexagonal lattice can be generated by moving two neighboring atoms along the two  vectors defining a triangular sublattice. This is the first distinctive characteristic that will be responsible for the exotic properties of the material. 

The dispersion relation of the Honeycomb lattice based on a simple tight binding calculation is known from the early works \cite{W47,SW58}. We will not repeat here the derivation which is very clearly written in any graphene review but will instead highlight the main properties. The first is that two atoms per unit cell implies a two dimensional  wave function to describe the electronic properties of the system. The entries of the wave function are attached to the probability amplitude for the electron to be in sublattice A or B. 

The nearest-neighbor tight binding approach reduces the problem to the
diagonalization of the one-electron Hamiltonian

\beq
{\cal H} =  -t \sum_{<i,j>} a^+ _i a_j
\label{oneehamil}
\eeq
where the sum is over pairs of nearest neighbors atoms $i,j$ on the lattice
and $a_i$, $a^+_j$ are the usual creation and annihilation operators.
The Bloch trial wave function has to be built as a superposition  of the atomic orbitals from
the two atoms forming the primitive cell:  
\beq
\Psi_{\bf k}({\bf r})=C_A\phi_A+C_B\phi_B.
\eeq 
The eigenfunctions and eigenvalues of the Hamiltonian are obtained from
the equation
\begin{eqnarray}
\left(
\begin{array}{cc} \epsilon & -t\sum_j e^{ia{\bf k·u_j}}
\\ -t\sum_j e^{ia{\bf k·v_j}} &\epsilon \end{array} \right)
\left( \begin{array}{cc} C_A \\ C_B \end{array} \right)
= E( {\bf k }) \left( \begin{array}{cc} C_A \\ C_B \end{array} \right)\;,
\label{eigen}
\end{eqnarray}
where ${\bf u_j}$ is a triad of vectors connecting an
A atom with its B nearest neighbors ($\delta_i$ in Fig. \ref{lattice}),
and ${\bf v_j}$ the triad of their respective opposites,
$a$ is the distance between carbon
atoms and $\epsilon$ is the 2$p_z$ energy level,
taken as the origin of the energy. The tight binding parameter $t$  estimated to be of the 
order of 3eV in graphene sets the bandwidth (6eV) and is 
a measure of the kinetic energy of the electrons. 
The eigenfunctions are determined by the the coefficients
$C_A$ and $C_B$ solutions of equation (\ref{eigen}). The
eigenvalues of the equation  give the energy levels whose dispersion relation is
\begin{align}
E({\bf k})=\pm t\sqrt{1+4\cos^2\frac{\sqrt{3}}{2}ak_x+
4\cos\frac{\sqrt{3}}{2}ak_x\cos\frac{3}{2}ak_y} \;,
\label{disprel}
\end{align}
in which the two signs define two energy bands which
are degenerate at the six vertices $ {\bf K}$   of the hexagonal Brillouin zone.
The dispersion relation is shown in the right hand side of Fig. \ref{lattice}.

The neutral system with one electron per lattice site is at half filling.  The Fermi surface consists of six Fermi points as can be seen in Fig. \ref{lattice} (only two are independent). This is the most important aspect of the system concerning its unusual properties. The existence of a finite Fermi surface (a Fermi line in two dimensions) in metals is at the hart of the Landau Fermi liquid standard model. It implies a finite density of states and the screening of the Coulomb interaction. Moreover it allows the construction of the Landau kinematics leading to the possibility of superconductivity and other collective excitations in the otherwise free electron system \cite{P93}.   

A continuum model can be defined for the low energy excitations 
around any of the Fermi points, say $K_1$,
by expanding the dispersion relation around it:
${\bf k}=K_1+{\bf \delta k}$ what gives from (\ref{eigen}) the Hamiltonian

\begin{eqnarray}
{\cal H}\sim -\frac{3}{2}ta \left(
\begin{array}{cc} 0 & \delta k_x+i\delta k_y
\\ \delta k_x-i\delta k_y  & 0 \end{array} \right)
.
\label{contin}
\end{eqnarray}

The limit $\lim_{a\to 0}{\cal H}/a$ defines the continuum Hamiltonian
\beq
H=v_F {\vec \sigma} .{\vec k},
\label{dirach}
\eeq
where $\sigma$ are the Pauli matrices and the parameter $v_F$ is the Fermi velocity of the
electrons estimated to be $v_F\sim 3ta\sim c/300$. Hence the low energy excitations of the system 
are massless, charged spinors in two spatial dimensions moving at a speed $v_F$. We must notice that the physical spin of the electrons have been neglected in the analysis, the spinorial nature of the wave function has its origin in the sublattice degrees of freedom and is called pseudospin in the graphene literature.  The same expansion around the other Fermi point gives rise to a time reversed Hamiltonian: ${\cal H}_2=v_F(-\sigma_x k_x+\sigma_y k_y)$. The degeneracy associated to the Fermi points (valleys in the semiconductor's language) is taken as a flavor. Together with the real spin the total degeneracy of the system is 4.

The chiral structure of the spectrum described above and the quantum mechanics nature of the condensed matter system (as opposite to QFT),  allows to test several predictions of the old relativistic quantum mechanics \cite{BD65}. 
In particular electrons in the graphene system will tunnel with transmission probability one through a step barrier hit at normal incidence. This phenomenon known  as the Klein paradox \cite{Ketal06,DC99} has been experimentally confirmed \cite{YK09,SHG09}. A similar phenomenon is the so--called Zitterbewegung or fast trembling motion of the electrons in external fields \cite{K06a,WZB07} whose experimental signature has been proposed in \cite{RZ09}. A particularly interesting phenomenon is the supercritical atomic collapse \cite{SKL07,PNC07} in graphene, a consequence of the large value of the ``graphene fine structure constant" to be discussed in Sect. \ref{sec_ren}. Another very interesting phenomenon is the possible realization of the Swinger mechanism \cite{S51}, i. e. the production of charged particle--antiparticle pairs driven by an external constant electric field \cite{AC08,DM10}.

The former phenomena are not only curious realizations of relativistic quantum mechanics, they have profound consequences on the experimental aspects of the material system most of them not observed before in  condensed matter.  In particular the Klein paradox implies that impurities and other most common sources of disorder will not scatter the electrons in graphene. This will evade the  Anderson localization  \cite{A58}, a very important result establishing that any amount of disorder in free electron systems in two space dimensions will localize the electrons turning the system to an insulator. The Klein tunneling is also partially responsible for the high mobility at room temperature and the excellent metallicity of the system. Zitterbewegung was suggested in \cite{K06a} as an explanation for the observed minimal conductivity of the samples \cite{G09}, one of the  most interesting aspects of graphene whose origin remains uncertain.
 
It can be proven that the Fermi points are topologically stable against deformations of the Hamiltonian preserving  the product of the discrete symmetries time reversal and spacial inversion \cite{MGV07}.  The topological stability is based on the non--trivial momentum space topology of the Fermi points \cite{Vol11}
and their associated non--trivial topological charge.
Moreover the wave function around the Fermi points have a non-trivial Berry phase that will be discussed in  Sect. \ref{sec_berry}.

\subsection{Multilayer graphene.}
\label{sec_bil}

\begin{figure}[!t]
\begin{center}
\includegraphics[width=4cm]{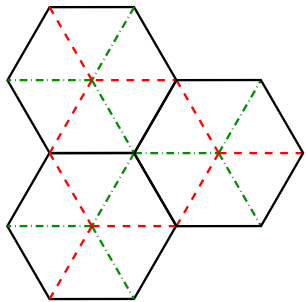}
\caption[fig]{\label{stacking} (Color online) Sketch of the three
possible positions of a given
  layer with respect to the others in a graphene stack. Bernal stacking
  ($1,2,1,2, \cdots$, is described by two inequivalent planes, while
  orthorhombic stacking, $1,2,3,1,2,3 \cdots$, requires the three
  inequivalent ones.}
\end{center}
\end{figure}

Multilayer  graphene consists in a superposition of several graphene layers interconnected by hopping terms. 
Depending on the relative orientation of the layers several stacking are possible. The two most common, 
staggered ($ABA$) and rhombohedral ($ABC$) are sketched in Fig.[\ref{stacking}].  The low energy limit of some multilayer systems give rise also to interesting QFT models defined in the continuum. Bilayer graphene is one of the best studied cases. It is relatively easy to obtain and manipulate experimentally and can be better than the monolayer for some potential applications \cite{NCetal06}. 

 We will denote the
two inequivalent atoms in layer $n$ as ($a_n,b_n$).
The simplest model introduces interlayer hoppings $t$ only between
nearest neighbors. The resulting hamiltonian for bilayer graphene in the vicinity of the
$K_1$ Fermi point  is
\beq
\label{bil}
\mathcal{H}(k)=\left(
\begin{array}{llll}
 0 & k^* & 0 & t \\
 k & 0 & 0 & 0 \\
 0 & 0 & 0 & k^* \\
 t & 0 & k & 0
\end{array}
\right)
\eeq
and the energy bands are
given by
\begin{equation}
\label{bila}
\epsilon_k =\pm \frac{t}{2} \pm
\sqrt{\frac{t^2}{4} + | \vec{k} |^2}.
\end{equation}
In the limit $E\ll t$, one can
obtain an effective hamiltonian for the lowest energy
bands~\cite{MF06}.
To this end, reorder the wavefunctions according $(a_1,b_1,a_2,b_2)\to (a_2,b_1,a_1,b_2)$,
so that in the new basis the hamiltonian becomes
\beq
\mathcal{H}(k)=\left(
\begin{array}{llll}
 0 & 0 & 0 & k^* \\
 0 & 0 & k & 0 \\
0 & k^* & 0 & t \\
 k & 0 & t & 0
\end{array}\right)\equiv\left(
\begin{array}{ll} H_{11} &H_{12}\\
 H_{21} &H_{22}
 \end{array}
\right)
\eeq
where $H_{ij}$ is a $2\times 2$ block.
The  identity
\begin{eqnarray}
& Det(\mathcal{H}-E)\\ \nonumber =
&Det\Bigl(H_{11}-H_{12}(H_{22}-E)^{-1}H_{21}-E\Bigr)\,Det(H_{22}-E)
\end{eqnarray}
shows that, for $E\ll t$, the substitution $H_{22}-E\to
H_{22}$ reduces the computation of the lowest energy bands to the
diagonalization of the $2\times 2$ effective hamiltonian
 \beq
 \label{bieff}
 \mathcal{H}^{eff}\equiv H_{11}-H_{12}H_{22}^{-1}H_{21}=-{1 \over t}\left(
\begin{array}{cc}
0 & k^{*2}\\
k^{2} & 0
 \end{array}
\right)
\eeq
This effective hamiltonian involves only the atoms $(a_2,b_1)$
which are not linked by $t$.  Under a QFT point of view the Hamiltonian is very exotic. It has chirality as the Dirac case but the dispersion relation is quadratic. Under the condensed matter point of view it is also a very interesting system where the Fermi surface reduces to a point but it has a finite density of states at the Fermi level similarly to standard metals.

 The former analysis can be easily generalized to multilayer graphene with
  rhombohedral stacking. This type of staking includes the links
  $(b_1-a_2, b_2-a_3,\ldots, b_{N-1}-a_N)$  and the effective hamiltonian,
  which  involves  only the unlinked  atoms $(a_1,b_N)$, is given by
\beq\label{muleff}
\mathcal{H}^{eff}=-{1 \over t^{N-1}}\left(
\begin{array}{cc}
0 & k^{*N}\\
k^{N} & 0
 \end{array}
 \right).
\eeq

It is easy to compute that there is a topological charge associated to this type of multilayer systems around the  Fermi points whose value   is $\pm N$ what also guarantees their topological stability \cite{MGV07}. Note however that there is no topological stability for the Bernal stacking. As we will see in Sect. \ref{sec_gapmulti} four Fermi interactions give rise to various spontaneous symmetry broken gapped phases in perturbation theory, something that does not happen in the monolayer system (discussed in Sect. \ref{sec_gapmono}).

\section{Quantum field theory aspects of graphene.}
\label{sec_QFT}

\subsection{Electron-electron interactions in monolayer graphene}
The effects of the electron-electron interactions in graphene has recently become a hot and highly debated topic. In the early times of the graphene boom, it was widely assumed that the main features of graphene could be reasonably studied using independent particle models. Theoretical work which did not include interactions where highly successful in explaining the anomalous features observed in the Integer Quantum Hall regime, the dependence on carrier concentration of the electron mobility, transport in p-n junctions (the Klein paradox) or the transparency at optical frequencies. The lack of interaction effects seemed confirmed by careful measurements of the electron compressibility, well explained by independent particle models. The advent of new samples with greatly enhanced homogeneity, and with carrier mobilities higher by about two orders of magnitude than the graphene flakes previously used, has changed completely the perception about the role of interactions.

The first hint was the observation of the Fractional Quantum Hall Effect in high mobility suspended samples. This phenomenon proved very elusive, but, when found, the associated energy gaps turned out to be higher than in other 2D electron liquids\cite{FQHE}\cite{DSetal09,BGetal09}.

\subsubsection{Long range Coulomb interactions. Graphene versus QED.}

The QFT modeling of Coulomb interactions in graphene has been explained in several review articles \cite{KN07,CGetal07,KUetal11,Voz11}. We will here summarize the main aspects of it.
 
As described in the previous section, the standard non-interacting model for the electronic excitations around  a single Fermi point in graphene   (in units $\hbar=1$) is given by
\begin{equation}
{\cal H}=  v_F \int d^2 {\bf r} \bar{\psi}({\bf r})
\gamma^i\partial_i \psi ({\bf r})\;, \label{freeH}
\end{equation}
where $i=1,2$, $\bar\psi({\bf r})=\psi^+({\bf r})\gamma^0$, and
the gamma matrices can be chosen as $\gamma_x=\sigma_2,
\gamma_y=-\sigma_1, \gamma^0=\sigma_3 $ . $\sigma_i$ are the Pauli matrices and
$v_{F}$ is the Fermi velocity, the only free parameter in the Hamiltonian.

The treatment of the Coulomb interaction is what makes a
difference between the graphene model and the usual QED(2+1). In
the definition of Coulomb interaction in lower dimensions we must
clarify weather we leave in flatland, i. e. the electromagnetic
field is really defined in two spacial dimensions, or the charges
are confined to a plane while the Coulomb field propagates in
three space dimensions. The scalar potential in classical
electromagnetism is defined by the Laplace equation in any number
of dimensions. Equivalently in QFT it is defined by the lagrangian
\begin{equation}
L=\int d^{(D+1)}x \quad {\cal F}_{\mu\nu}{\cal F}^{\mu\nu}
\;\;\;,\;\;\;{\cal F}_{\mu\nu}=\partial_\mu A_\nu - \partial\nu
A_\mu \;, \label{gaugeaction}
\end{equation}
which has two space derivatives. In both cases the effective
potential in Fourier space is $V(k)\sim 1/k^2$. To see the spacial
dependence of the interaction we must Fourier transform $V({\bf
k})$ and obtain the know result
\begin{equation}
V({\bf r-r'}) \sim
\left\{ \begin{array}{lrlr} \frac{1}{\vert {\bf r-r'}\vert} &{\rm D=3} \\
\log(\vert {\bf r-r'}\vert ) &{\rm {\rm D=2}} \\
\vert {\bf r-r'}\vert &{\rm D=1} 
\end{array} \right\} 
\label{coulombD}
\end{equation}
In two spacial dimensions the Coulomb interaction in real space is
logarithmic. In the real world only the charges are confined to
the graphene plane while the interaction propagates in three
dimensions. This is encoded in the  usual form of the
(instantaneous) Coulomb interaction:
\begin{equation}
{\cal H}_{int} = \frac{e^2}{2 \pi} \int d^2 {\bf r_1} \int d^2
{\bf r_2} \frac{\bar{\Psi} ( \bf{r_1} ) \Psi ( \bf{r_1} )
\bar{\Psi} (\bf{r_2} ) \Psi ( \bf{r_2} )} {| \bf{r}_1 - \bf{r}_2
|}.
 \label{coulomb}
\end{equation}
The interaction (\ref{coulomb}) is inadmissible in a QFT approach.
It is not only non-relativistic (action-at-a-distance) but it is
also non-local. We can try to formulate the interacting problem in
a QFT language by coupling the fermions to a gauge field through
the minimal coupling prescription
\begin{equation}
L_{int}=g\int d^2x dt j^\mu (x,t)A_\mu (x,t)\;,
\end{equation}
where $g=e^2/4\pi v_F$ is the dimensionless coupling constant, and
the electronic current is defined as
$$j^\mu =(\overline\Psi\gamma^0\Psi,v_F \overline\Psi\gamma^i\Psi)\;.$$
Now we have to face the problem of coupling the two-dimensional
current to a three dimensional gauge potential. In ref.
\cite{GGV94}  this difficulty was solved by using the Feynman
gauge
$$\nabla_\mu A^\mu=0.$$

The full Hamiltonian looks like that of (non-relativistic) quantum
electrodynamics in two spacial dimensions:
\begin{equation}
{\cal H}= \hbar v_F \int d^2 {\bf r} \bar{\Psi}({\bf r})
\gamma^\mu(\partial_\mu -igA_\mu) \Psi ({\bf r})\;. \label{fullH}
\end{equation}
Notice that non-relativistic can be used in two different ways.
One is  non-covariant in the sense that the time and space
components of the current have different velocities. This lack of
covariance does not invalidate most of the rules that make life
easy in quantum field theory  as the Ward identities. A more
severe form of non-relativistic is setting the rate $v_F/c=0$.
This amounts to considering that there is no time dependence in
the gauge propagator. The results obtained in this case can
invalidate some of the QFT results as we will see. In ref.
\cite{GGV94} it was shown using renormalization group methods that
the hamiltonian (\ref{fullH}) runs to an infrared fixed point
where the Fermi velocity equals the speed of life recovering the
full Lorentz invariance.

In any case, despite the formal identification of the hamiltonian
(\ref{fullH}) with QED(2+1), the different photon propagators
produce different results even if we adopt a relativistic approach
(i.e. assume a "retarded" Coulomb interaction) from the beginning.

\subsubsection{Perturbative renormalization.}
\label{sec_ren}

Ultraviolet divergences arise in quantum field theories due to the
singular behavior of the fields at very short distances in real
space - or at very large energies in Fourier space-. The existence
of an underlying lattice in most condensed matter theories
provides a natural cutoff and the issue of divergences can usually
be ignored. Infrared divergences appear also in massless theories
-and hence there are also present in the graphene - and can be
treated  with different techniques. Here again
condensed matter resorts to the  finite size of the samples to
avoid them. Renormalization \cite{N78,C84} is a prescription to
get rid of ultraviolet divergences and construct sensible models
where physical quantities can be accurately computed.  The idea is
that the ultraviolet divergences can be cancelled by a
redefinition of the parameters (mass, coupling constant, wave
function) of the theory by adding  counter terms to the
lagrangian. The process is usually done order by order in
perturbation theory. If done appropriately, at each order one
finds finite results {\it independent of the cutoff} in the
computation of physical observables.

Depending on the number and badness of the infinities appearing in
the computation of the Feynman diagrams quantum field theories are
classified as nonrenormalizable, strictly renormalizable, and
superrenormalizable. In most cases renormalizable models have an
infinite number of divergent graphs -as we go to higher loops in
the computation- but only a finite "type" of divergences given by
the so-called primitively divergent graphs. The models describing
strong and electroweak interactions in three space dimensions are
strictly renormalizable what means that the number of primitively
divergent graphs exactly equals the number of free parameters in
the model. In this case at higher loop order we get higher powers
of the divergent logarithms and it is possible to perform a formal
sum of the perturbative series for the parameters.

Lowering the dimension of the space, strictly renormalizable
theories as QED become superrenormalizable: only a finite set of
graphs need overall counterterms. The number of divergent graphs
will still be infinite as if there is a one-loop graph G
diverging, any graph containing G as a subgraph will also diverge.
But if we renormalize G by adding a counterterm to it, all these
higher order graphs will automatically be finite. The usual rule
that assigns higher log powers to higher order in perturbation
theory simply does not work and there is no problem with it.

In the case of having a gauge symmetry as in QED, the process of
renormalization can interfere with gauge invariance and more care
is needed to complete the job. If we maintain the postulates of
QFT, namely unitarity, locality and Lorentz invariance, there are
Ward identities that relate some of the renormalization functions.
If some of the postulates are broken as usually happens in the
condensed matter applications, some identities might not work.

\begin{figure}
\begin{center}
\includegraphics[scale=0.5]{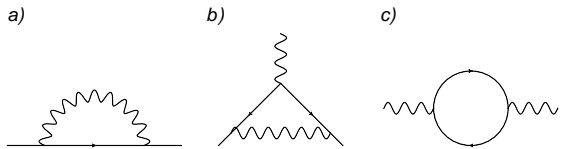}
\caption{Primitively divergent Feynman graphs in QED(4). (a) Fermion self--energy,  (b) vertex, (c) photon self-energy.}
\label{primitive}
\end{center}
\end{figure}

Standard QED(3+1) is a strictly renormalizable theory. The coupling constant is dimensionless and, in the massless case, the model is scale invariant.  The three primitively divergent diagrams shown in Fig. \ref{primitive} are associated to the three free parameters in the theory (the coupling constant and the electron and photon wave functions).
The theory is strictly renormalizable and all divergences at any order in perturbation theory can be cured by a proper redefinition of the parameters.  QED(2+1) is  super-renormalizable. The coupling constant has dimensions of energy what improves the convergence of the perturbative series. Massless QED (2+1) is ultraviolet finite although it has infrared divergences \cite{MRS05}. Graphene sits in between  QED(3) and QED(4).  Of the three graphs shown in Fig. \ref{primitive} only the electron self--energy diverges and in the case of considering a static photon propagator only the spatial part has a logarithmic divergence leading to the renormalization of the velocity parameter in (\ref{fullL}).
In the static model widely used in condensed matter applications, the electron and photon propagators in momentum space are given by
\begin{equation}
G_0(k^0, {\bf k})=i\frac{\gamma^0 k_0+ v\bm{\gamma}
\cdot {\bf k}}{-(k^0)^2+v^2{\bf k}^2}\;,
\label{elprop}
\end{equation}
\begin{equation}
\label{barephoton}
\Pi_0( {\bf k})=\frac{1}{2}\frac{1}{\vert {\bf k}\vert}\;.
\end{equation}

The renormalization functions associated to the electron self-energy are defined as:
\begin{equation}
G_0^{-1}-\Sigma(k^0,{\bf k})=Z_\psi^{-1/2}(k^0,{\bf
k})[k^0\gamma^0-Z_v(k^0,{\bf k})v\bm{\gamma} \cdot {\bf k}],
\label{zelectron}
\end{equation}
The extra parameter $v$ appearing in (\ref{elprop}) in the graphene case has an associated renormalization function $Z_{v}$ which is a new feature characteristic of graphene. In the Lorentz invariant massless QED in any dimensions, only the electron wave function renormalization is associated to the the electron self energy. 

The wave function renormalization
\beq
Z_\Psi\sim\frac{\partial \Sigma(\omega, {\bf
k})}{\partial\omega}\vert_{\omega=0}\;,
\eeq
defines the anomalous
dimension of the field 
\beq
\gamma=\partial \log Z_\Psi/\partial l,
\eeq
($l$ is the RG parameter that for a hard cutoff $\Lambda$ is $l=\log\Lambda$) and, hence the asymptotic behavior of the fermion propagator:
\beq
G(\omega, {\bf k})\sim_{\omega\to 0}\frac{1}{\omega^{1-\eta}}\;.
\eeq
The function $Z_v$ defines the running of the coupling v whose beta function is defined as
\beq
\beta_v=\partial Z_v/\partial l.
\eeq
The fixed points of the system are determined by the zeros of this beta function and the running of the velocity is given by the local behavior of the beta function around the fixed point \cite{C84}.

It can be  seen that in the static model the perturbative series is organized in terms of a coupling constant
%\beq
$\alpha_G=\frac{e^2}{4\pi v_F}.$
%\label{gbare}
%\eeq
Plugging-in the Fermi velocity  measured in nanotube experiments  $v_F\sim c/300$ \cite{LB01}, and the bare electron charge,  we get $\alpha_G\sim 2.3 - 2.5$. Although this value puts the model in the strong coupling regime, the results obtained in perturbation theory agree with the experimental findings. A possible explanation lies on the fact that the effective constant in a material has an extra parameter: the dielectric constant taken as one in the vacuum. In samples on a substrate the dielectric constant reduces the effective coupling constant:
% \beq
$\alpha_G=\frac{e^2}{4\pi \varepsilon_G v_F},$
%\label{gepsilon}
%\eeq
where $\varepsilon_G$ includes intrinsic contributions and effects due to the environment in which graphene is immersed.  A simple estimate gives $\varepsilon_G = ( \varepsilon_{air} + \varepsilon_{subs} ) / 2 \sim 2-3$ for typical substrates. The actual value of the intrinsic contribution to the dielectric constant due to the graphene layer itself is still object of vivid controversies. By extrapolating measurements of the excitation spectra in graphite a large intrinsic dielectric constant in single layer graphene of the order of $ \varepsilon_G\sim 13$ was proposed in \cite{RU10}. An independent estimate of $\alpha$ in graphene has been obtained from measurements of the carrier-plasmon interaction in samples with a finite carrier concentration\cite{BSetal10}. The result, $\alpha \approx 2.2$ is consistent with  previous analysis \cite{A06} and from recent numerical calculations \cite{WSetal11}. Theoretical values of the order of $ \varepsilon_G\sim 4$ have also been reported using various approximations \cite{HD07,WK11}.

\subsubsection{Velocity renormalization.}
\label{sec_velren}
The computation of the graph in Fig. \ref{primitive} (a) gives
\beq
\Sigma^{(1)}(\mathbf{k})\sim \frac{g}{4}v\bm{\gamma}\cdot\mathbf{k}
\log \frac{{\bf k}^2}{\Lambda^2},
\label{cutoff}
\eeq
where $\Lambda$ is a high energy cutoff and g is the bare coupling. The logarithmic divergence can be fixed by redefining the fermion velocity $v$ which becomes, after finishing the renormalization procedure, energy-dependent.
The onle loop renormalization group equation can be solved to obtain
\beq
\frac{v_F(E)}{v_F(E_0)}=1-\frac{g(E_0)}{16\pi} \log(E/E_0),
\label{oneloopveloc}
\eeq
which relates the value of the Fermi velocity at an energy $E$, with that at a reference energy $E_0$, assuming that the two energies are sufficiently closed. $E_0$ is not a cutoff of the order of
the bandwidth but any energy where the Fermi velocity is determined by experiments. The corresponding diagram in massless QED(3) does not have a logarithmic divergence because the photon propagator has an extra inverse power of momentum. In QED(4) this diagram induces a wave function renormalization. It is interesting to note that the electron charge in the graphene model is not renormalized (after fixing  the velocity divergences, the photon self-energy is finite at all orders in perturbation theory)  but the velocity renormalization induces a renormalization of the graphene structure constant similar as the one obtained in QED(4). This issue will be explored further when discussing the experimental observations in Section \ref{sec_expv}.

Since the bare coupling constant of graphene can be large at the experimentally accessible energies, the former renormalization scheme can be improved by performing a 1/N expansion \cite{H74} where $N$ is the number of fermionic species. In the case of graphene with the spin and Fermi points degeneracies the physical value is  $N=4$. This procedure was followed in the early publications \cite{GGV96,GGV99}  and was later retaken in \cite{FA08,DL09b,GGG10}.

The simplest non-perturbative calculation amounts to compute  the electron self--energy graph in Fig.~\ref{primitive}(a) with an effective  propagator obtained from  the  resummation of the planar diagrams dominant  in the 1/N approximation. The $\Pi_{00}$ component of the one loop photon self--energy in Fig.~\ref{primitive}(b) is given by
\beq
\Pi({\bf k}, \omega)=\frac{e^2}{8}\frac{{\bf
k}^2}{\sqrt{v_F^2{\bf k}^2-\omega^2}}\;, \label{bubble}
\eeq
hence, the effective Coulomb potential obtained by summing the geometric Dyson series for the photon self--energy in the so--called random phase approximation (RPA)  is
\beq
\Pi(\omega, {\bf k})=\frac{-i}{2\vert{\bf k}\vert+
\frac{e^2}{8}\frac{k^2}{\sqrt{v_F^2 k^2 - \omega_k^2}}}.
\eeq
The electron self--energy computed with this expression gives the following  $\beta$ function for the Fermi velocity \cite{GGV99,FA08}:
\beq
\beta_v=-\frac{8}{\pi^2 N}\big(1+\frac{\arccos g}{g\sqrt{1-g^2}}\big)
+\frac{4}{\pi g}, \label{large_n}
\eeq
where $l$ is the scaling parameter ($l=\log\Lambda$ in a cutoff scheme), and $g=\frac{Ne^2}{32 v_F}$. The large N limit amounts to take the limit $N\to\infty$ keeping $g$ fixed. As we see the dependence on $g$ is non-perturbative and the growth of the velocity at low energies is slightly different than that in (\ref{oneloopveloc}).

A remark is here in order. In graphene many classes of lattice defects can be
described by gauge fields coupled to the two dimensional Dirac
equation \cite{VKG10}. The standard techniques of disordered
electrons \cite{AS06} can be applied to  graphene by
averaging over the random effective gauge fields. A random distribution of defects leads to a random gauge
field, with variance related to the type of defect and its
concentration. These random fields when
treated perturbatively, lead to a renormalization of the  Fermi velocity which 
makes it to decrease at lower energies opposing the upward renormalization induced
by the long range Coulomb interaction. The simultaneous presence of interaction and disorder
gives rise to new interesting fixed points. The issue was analyzed in \cite{SGV05}. The most interesting case arises when considering a random gauge potential which models elastic distortions and some topological defects.  There is  a line of fixed points with 
Luttinger-like behavior for each disorder correlation strength
$\Delta$ given by $v_F^*=2e^2/\Delta$. An extensive analysis of the issue in the large N limit is done in \cite{FA08}.

Another interesting prediction done within this framework is the linear dependence of the quasiparticles lifetime  with the energy \cite{GGV96}, a distinctive of the marginal Fermi liquid behavior \cite{Vetal89}. Experimental evidences  of this behavior have  been described in \cite{Zetal06,Jetal07,LLA09,KLetal11}.

\subsubsection{Experimental confirmation of the Fermi velocity renormalization.}
\label{sec_expv}

\begin{figure}
\begin{center}
\includegraphics[scale=0.6]{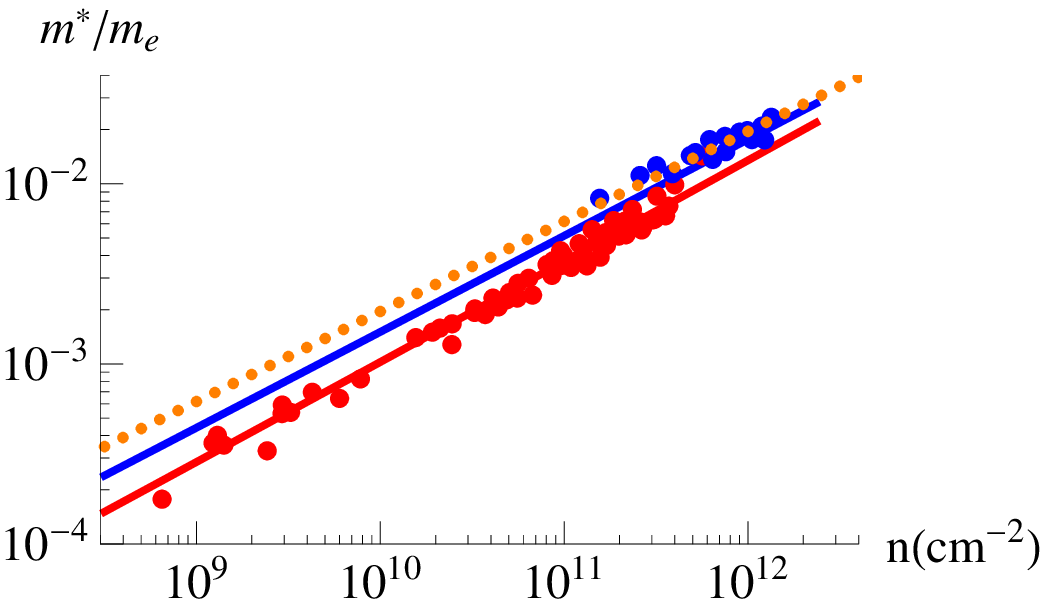}
\hspace{0.5cm}
\includegraphics[width=8cm]{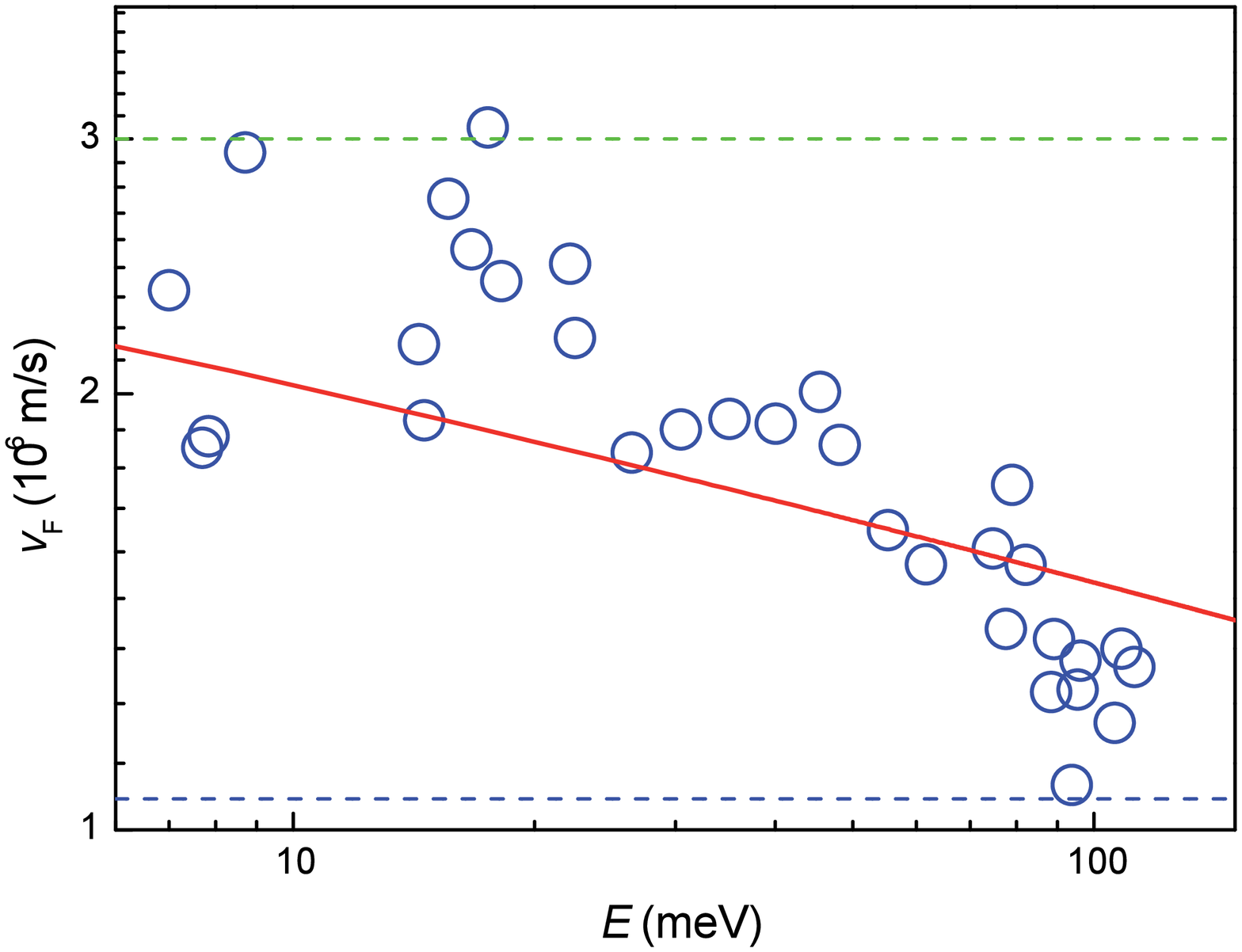}
\caption{Left: Experimental measurements of the effective mass of carriers in high mobility graphene samples. Red dots give results in suspended samples, and blue dots are for samples on BN substrates. The red and blue lines are fits using eq. \ref{large_n}. Red: $\varepsilon_G = 1$. Blue: $\varepsilon_G = 4$. The dotted orange line shows a fit obtained neglecting the renormalization of the Fermi velocity. Right: Plot of the Fermi velocity renormalization from the measurements done in \cite{EGetal11}.}
    \label{fig_velocity}
\end{center}
\end{figure}

The RG prediction of the velocity renormalization has been  very recent confirmed in several experimental reports  \cite{EGetal11,LLA11,ZSetal11} following earlier more indirect evidences
\cite{BOetal07,Letal08}. The experimental evidence more directly related to  the physics discussed in this review is that in \cite{EGetal11}.  The experiment measures the effective mass of graphene at different carrier densities in  high mobility suspended samples and also in graphene on a BN substrate. The experiment probes the suspended graphene samples at  low energies in a range form 100meV  to 0.2 meV never reached before. 
They use the temperature dependence of Shubnikov-de Haas oscillations to infer the dependence of the Fermi energy, $\epsilon_F$, on the area of the Fermi surface, $S_F$. The effective mass is defined as
\beq
m_{eff} = \frac{\hbar^2}{2 \pi} \frac{d S_F}{d \epsilon_F}
\eeq
so that for graphene
\beq
m_{eff} = \frac{\hbar k_F}{v_F}
\eeq
A comparison of experimental results an fits based in eq.~(\ref{large_n}) are shown in the left hand side of Fig.~\ref{fig_velocity}.

The behavior of the inverse running coupling constant in graphene taken from the data in \cite{EGetal11}
is shown in the right hand side of Fig. \ref{fig_velocity}.

The difficulty of achieving these results is similar to that of going several orders of magnitude higher in energy to confirm the fine structure constant renormalization \cite{Voz11}. It is interesting to compare with the case of QED(3+1). The fine structure constant is defined by 
$\alpha_{QED}=e^2/(4\pi\epsilon_0 c)$ where $e$ is the electron charge, c the speed of light, and $\epsilon_0$ the dielectric constant of the vacuum. As such, it is a dimensionless quantity whose universally known value is $\alpha^{-1}=137$. From its definition, measuring the fine structure constant amounts to measuring the electron charge with the higher possible accuracy since the other quantities in the definition are real constants \footnote{We will stay on the side of c being a universal constant and wait to  see what happens with the OPERA measurements \cite{Op11}.}.

While the value of $e$ obtained through Hall resistance measurements reached an accuracy of six significant digits \cite{KDP80}, the actual value got from the magnetic moment of the electron has an accuracy of 0.7 parts per billion \cite{HFG08}. This allows a determination of $\alpha_{QED}$ of similar accuracy. The  calculation includes contributions from 891 Feynman diagrams and is one of the most demanding comparisons of any calculation and experiment ever performed \cite{GHetal08}. The impressive accuracy reached at the solid state energies (around 100mK considered as zero for the high energy running coupling constant) is spoiled at energies of the order of the proton mass (1GeV) where strong interaction diagrams contribute to the determination of $\alpha_{QED}$.  A relatively recent measurement done at the Large
ElectronÐPositron collider (LEP) at CERN provides a value of $\alpha^{-1}(M_Z^2)=128.946 ± 0.030$ \cite{L300,Er99}. From this we can see that the slope of the variation in two order of magnitude is similar in the two cases.

\subsubsection{Retarded Coulomb interactions and the structure of the perturbative series.}
\label{sec_ret}

The  success of the effective description of the long range Coulomb interaction in graphene by means of a static Coulomb potential described before that lead to the observation of the Fermi velocity renormalization  means that retardation effects of the order of $v_F/c\sim 1/300$ are ineffective in the today's experimental accessible region. Nevertheless there are important issues concerning the structure of the QFT model of graphene that we will address here. 

First the inclusion of a covariant photon propagator 
\begin{equation}
\label{retphoton}
\Pi_0(\omega,{\bf k})=\frac{1}{2}\frac{1}{\sqrt{\omega^2-\vert {\bf k}\vert^2}},
\end{equation}
instead of (\ref{barephoton}) leads to important conceptual differences in the infrared behavior of the model. 

The one loop computation of the electron propagator in this case gives the result \cite{GGV94}:
\beq
\beta_v=\frac{1}{v}\frac{1-2v^2+4v^4}{(1-v^2)^{3/2}}
\arccos v+\frac{1-4v^2}{1-v^2},
\label{betav}
\eeq
where the speed of light has been put to one.
The equation $\beta_v=0$ has the solution $v=1$ which in our units means that we have a fixed point of RG where the electronic velocity equals the speed of light.  

Since the electric charge $e$ is not renormalized, the value of the coupling constant at the fixed point $g^*$ is the fine structure constant of QED: 
\beq
g^*=\alpha_{QED}=\frac{1}{137}.
\eeq

This issue was initially studied in \cite{GGV94} was later confirmed in a non-perturbative approach in \cite{GMP10b}. Moreover there is a wave function renormalization at the one loop level that has been may give rise to a non--trivial critical exponent.

\subsection{ Chiral symmetry breaking or gap opening in monolayer graphene. }
\label{sec_gapmono}
Opening a gap (or generating a mass) in graphene is crucial for the electronic applications. Under the QFT point of view, the electron mass (gap) is protected by the 3D version of chiral symmetry
and hence  a gap will not open by radiative corrections
at any order in perturbation theory. As discussed in \cite{MGV07} interactions or disorder can open a gap in the sample provided that the product of time reversal times inversion  symmetry is broken (staggered  potential, external magnetic field), or when the two Fermi points are involved (Kekul'e distortion). A discussion of the various "masses" that can be generated in graphene when the valley and spin degrees of freedom are taken into account will be done in Sect. \ref{sec_topomono} when discussing the topological insulator aspects of graphene. Even if the two Fermi points are taken into account in a four dimensional formalism, that a gap will be spontaneously generated by interactions in monolayer graphene is very unlikely and there are so far no experimental evidences for it.

\subsubsection{Long range Coulomb interactions.}
\label{sec_gaplong}
One of the interesting features of QED(3) is that a fermion mass can
be generated dynamically, breaking chiral symmetry. Starting with
massless bare fermions, they can acquire a dynamical mass through
nonperturbative effects. Writing the full fermion propagator as
$$S^{-1}(p)=A(p){\bf \gamma}.{\bf p}+B(p),$$
 a nonzero solution for B(p) implies a
nonzero condensate, and signals dynamical mass generation. The
infrared value of the dynamical mass function defined by
$$m(p)=\frac{B(p)}{A(p)},$$ $m(0)=B(0)/A(0)$ can also be used as order
parameter.

The standard
non-perturbative approach to study dynamical fermion
mass generation is to solve the Dyson-Schwinger (DS) equation:
\beq
G^{-1}(p)=G^{-1}_0(p)-\int\frac{d^3 k}{(2\pi)^3}\gamma_\mu\Gamma_\nu(p, k)D_{\mu\nu}(p-k),
\label{eq_SD}
\eeq
where $\Gamma_\nu(p, k)$ is the vertex function and $D_{\mu\nu}$ the photon propagator. Eq. (\ref{eq_SD}) can be decomposed in a couple of equations for $A(k^2)$ and  $B(k^2)$:
\beq
A(p^2)=1+\frac{1}{4p^2}\int \frac{d^3 k}{(2\pi)^3} tr[\gamma . k \gamma_\mu G(k)\Gamma_\nu D_{\mu\nu}(p-k)].
\eeq
\beq
B(p^2)=-\frac{1}{4}\int \frac{d^3 k}{(2\pi)^3} tr [\gamma_\mu G(k)\Gamma_\nu D_{\mu\nu}(p-k)].
\eeq
If the DS equation for $B(p2)$ has only vanishing solution,
the fermions remain massless and are stable against gauge fluctuations.

The issue of chiral symmetry breaking (CSB) in massless fermion models in QFT is far from being settled. Lower dimensional QED, in particular QED(3) was proposed in the early times as a toy model for CSB \cite{P84,AP81,JT81,N89} and confinement \cite{Po75,Po77}. It is clear that no mass will be generated in the weak coupling limit. But, as we described in Sect. \ref{sec_ren} the bare coupling constant of graphene is not small. 

In the strong coupling regime the best studied approximation is based on a large number of fermionic species,  the $1/N_f$ expansion (not to be confused with the $1/N$ methods used in QCD where $N$ is the number of colors of an $SU(N)$ gauge theory). The crucial difference between the perturbative
and non-perturbative approaches is that the DS
equation is nonlinear which makes it possible for chiral
phase transition to happen at the bifurcation point. Spontaneous chiral symmetry breaking in QED(3) in the 1/N approximation usually requires an unphysical number of fermionic species $N<2-3.3$ \cite{ABetal86}. The situation with graphene is even worse since there is no mass parameter to begin with. The gauge propagator is crucial in these approaches and hence it is important to distinguish between QED(3) and graphene. This problem has been addressed in \cite{GGM01} with the conclusion that for small $g$ there is no solution with spontaneous chiral symmetry breaking. A very good updated review with a fair list of references can be found in \cite{Se11}.

The simplest non-perturbative calculation that can be done to study the issue of the gap opening in graphene is the RPA type  described in Sect. \ref{sec_velren} \cite{GGV99}. The absence of a constant term in the inverse electron propagator ensures that no mass is generated in this approximation. The 1/N expansion approach to the problem  has been revised recently in \cite{G10} with the conclusion that  a gap will not open for the physical values of the electronic degeneracy in graphene (N=4). A variational approach to the excitonic phase transition in graphene including the renormalization of the Fermi velocity \cite{SSG10} also produces quite negative results. At finite temperature the critical $N_c$ is temperature dependent \cite{KK93} while a finite chemical potential leads to strong suppression of the critical fermion flavor Nc and the dynamical fermion mass in the symmetry broken phase \cite{LL10}. Volume effects have been analyzed in \cite{GFW08} trying to explain the discrepancies between the continuum and lattice results.

The situation can be improved by the presence of an external magnetic field given rise to the so--called  the magnetic catalysis\cite{GMS94}. This possibility was put forward in the early papers \cite{K01,K01b,GGetal02} and is discussed at length in the review \cite{GSC07}.

\subsubsection{Four fermi local interactions.}
\label{sec_gapshort}
Short range interactions, such an onsite
Hubbard term $U$ are irrelevant in an RG sense \cite{Voz11} in the weak limit. This is due to the vanishing density of states at the Fermi level that occurs at Fermi points, a characteristic of graphene. This is a very important aspect since short range interactions are responsible of the Fermi liquid properties of usual metals (having a finite Fermi surface) in two space dimensions. Nevertheless the density of states at low
energies can be  increased by the presence of disorder and a finite temperature what in turn,
enhances the effect of short range interactions
\cite{GGV01}. These interactions can be relevant in the strong coupling regime of the hexagonal electronic and optical lattices \cite{ST92,Petal05}.

Four Fermi interactions were considered in low dimensional models in high energy physics again searching for chiral symmetry breaking and confinement. One of the most popular models is the Nambu-Jona Lasinio (NJL) in two space dimensions \cite{NJL61}. The original Nambu-Jona-Lasinio model was an extension of the BCS-theory of superconductivity to the domain of spontaneous symmetry breaking in the strong interaction, one of the best examples of hybridization between condensed matter and high energy physics. The lagrangian density is given by
\beq
L=-i \bar\Psi \gamma^\mu \partial_\mu +g\left[(\bar\Psi\Psi)^2 - (\bar\Psi\gamma_5\Psi)^2\right],
\label{eq_NJL}
\eeq
The four-fermion interaction is attractive for fermions and antifermions of opposite chirality. Fermion-antifermion pairs form bound states, which form a condensate:
$$<\bar\Psi \Psi> \neq 0$$
that changes the nature of the vacuum of the theory.

In the condensed matter language the NJL model is similar to the Hubbard model that has been extensively studied in lattice models. This type of  interactions has been used in the 
Honeycomb lattice searching for the possibility of gap opening in \cite{He06,S07,SS07,DL09a,DL09b,JHS09}. More recently lattice gauge field theory techniques are being used to explore this possibility. An interesting recent analysis of the possible phases arisen in the Jona-Lasinio model in three dimensions is done in \cite{GJ10}. Non-perturbative constructive field theories techniques have also been used and are described   in \cite{GMP11}. Although the issue remains controversial, it is fair to say that there is no conclusive evidence up to now that the physical parameters of graphene lie in the region where chiral symmetry breaking will occur.

A gap can open in the sample by various extrinsic means although without altering the basic properties of graphene, the produced gaps are very small and not entirely controllable. 
Small gaps have been experimentally reported in samples on a substrate whose lattice is commensurate with that of graphene \cite{GFetal07}. Graphane, a hydrogenated compound having hydrogen atoms forming $sp^3$ bonds has a reasonable gap and is an interesting material on its own grounds \cite{EGetal09}.

\subsection{Electronic interactions in multilayer systems.}
\label{sec_gapmulti}

The situation concerning electron--electron interactions is much more interesting in bilayer graphene. The quadratic dispersion relation and a finite density os states at the Fermi surface enhances the role of four Fermi interaction  and a variety of broken symmetry phases can arise similarly to what happens in the  one dimensional case \cite{Hal81}.

New experiments in high mobility samples hint to the existence of novel phases at zero carrier concentration and low temperatures. The first indication came from the electron compressibility measurements reported in\cite{FMY09}. These experiments showed a rise in resistivity near the neutrality point, consistent with a tendency towards an insulating state, although the resistivity never went above a value of a few thousand ohms. The existence of a gapped insulating phase was reinforced by extrapolations from measurements made at finite magnetic fields\cite{Metal10}, which suggest a gap of about $\Delta \approx 2$ meV.

The experiments reported in\cite{Metal11} present a different picture.  The density of states at the Fermi energy is inferred from careful measurements of the carrier density and temperature dependence of the resistivity in a number of high mobility suspended samples. The results show a crossover to a low temperature regime where the density of states is significantly reduced. The band structure of bilayer graphene changes from parabolic to four Dirac cones, due to trigonal warping. The crossover found experimentally occurs at an energy of about 6 meV, which is substantially larger than the crossover energy related to trigonal warping, abou1 1 meV. Hence, the results suggest a spontaneous symmetry breaking associated to interactions. This explanation is supported by measurements performed at finite magnetic fields, which imply that the lowest Landau level is fourfold degenerate, while independent electron calculations give an eightfold degenerate $n=0$ Landau level. In any case, the results reported in\cite{Metal11} do not show a finite gap at any concentration.

A last batch of recent experiments is discussed in\cite{Vetal11}. The conductivity of s suspended sample at the neutrality point was measured as function of bias voltage, magnetic field, and perpendicular electric field. The results indicate a gap of for bias voltages between abut -3 and 3 meV at zero magnetic field, zero electric field and zero carrier concentration. The differential conductance shows peaks at voltages above and below this gap, while they tend towards a constant value at larger bias voltages. Peaks adjacent to a gap are a hallmark of tunneling experiments into superconductors, although there is no apparent reason for these peaks to show up in a d. c. transport experiment.

All the experiments described above coincide in that a broken symmetry phase due to electron electron interactions seems likely in bilayer graphene near the neutrality point and in the absence of a perpendicular electric field. They differ in many details, however, and the experiments in\cite{Metal11} suggest a gapless phase, while those in\cite{FMY09}, and especially those in\cite{Vetal11} seem to imply the existence of a gap. The experiments analyze suspended samples with high electron mobility, although the mobilities differ, being highest in\cite{Metal11}, and lowest in\cite{FMY09,Metal10}, while the samples in\cite{Vetal11} show an intermediate value. The experimental setups also differ: the measurement in\cite{FMY09,Metal10} give the values of the compressibility in samples with one back gate, while the experiments in\cite{Metal11} and in\cite{Vetal11}  report the d.c. conductivity in samples with no and with two gates respectively.

On the theory side, it was soon realized that the combination of parabolic bands and short range, screened interactions lead to logarithmically divergent susceptibilities. The variety of electronic degrees of freedom in bilayer graphene allows for many possible broken symmetry ground states, which, in turn, can be induced by appropriately tuned interactions\cite{NNPG06,MAF07,HBPM08,BY09,LATF10,VY10,NL10,ZMPM10,V11}. There is a gapless, nematic phase consistent with the experiments in\cite{Metal11}, and different gapped phases could explain the results in\cite{FMY09}, while the interpretation of the experiments in\cite{Vetal11} remains less clear. Some of the proposed phases break time reversal symmetry, leading to states with similar properties to a 2D electron gas in the Integer Quantum Hall regime\cite{H88}. Recent calculations suggest that many possible phases are almost degenerate, with energy differences per atom below 1 meV\cite{JZM11}. 

Interactions in a graphene bilayer are screened, do that they decay faster than the inverse of the separation beyond some distance. The nature of the broken symmetry phase may depend on the value of the screening length\cite{TV11}. This leads to the intriguing possibility that the two phases which look more stable, a gapped layer antiferromagnet and a metallic nematic phase, could have been observed in different experiments, depending on the number and position of gates.
Further complications are introduced by strains, which are probably unavoidable in suspended samples. Disorder can induce, in some circumstances, local gaps\cite{LHS11}. This situation is reminiscent of other materials with many competing interactions, such as the cuprate superconductors or the manganites, where the interpretation of the low temperature phases is still debated. At least, graphene has a simple, stochiometric composition.

\section{Gauge fields from lattice deformations.}
\label{sec_elastic}
One of the most fascinating facts about graphene is the possibility to find experimental realizations quite abstract theoretical ideas. There is a particular aspect that resembles what happened in the physics of the standard model in the last century: Gauge bosons were postulated on theoretical necessities and found experimentally afterwards \cite{}. The emergence of gauge fields associated to lattice deformation has been discussed at length in ref. \cite{VKG10}. We will here summarize the main aspects before revising the experimental observation of them that has taken place in the last year.  As described in \cite{VKG10} gauge fields have emerged from three different sources: Topological defects, elastic distortions, and from a covariant, geometrical formulation. The physical  origin of them is rooted in the tight relation between the spinorial nature of the carriers in graphene and the lattice structure. As described in the introduction the spinor degree of freedom is associated with the non trivial geometry of the Honeycomb lattice.

ARPES measurements show that, despite significant deviation from planarity of the crystal, the electronic structure of exfoliated suspended graphene is nearly that of ideal, undoped graphene \cite{KLetal11}.

%\subsection{ Gauge fields from topological defects.}

%Fullerenes are macromolecules of carbon atoms forming closed structures made of hexagons and topological defects (pentagons).  Among them the more common is the $C_{60}$ which has the structure of a truncated icosahedron (see Fig. \ref{fig_c60}). According to the Euler theorem the number of polygons needed   to form a closed polyhedron is fixed by the Euler characteristic $\chi$ is defined for the surfaces of polyhedra in terms of the number of vertices (V), edges (E) and faces (F) by $\chi$=V-E+F. Any convex polyhedron's surface has Euler characteristic $\chi$=2 which is the number corresponding to the sphere. An Euler theorem establishes that for any polyhedron with three edges emanating from each vertex we have $\sum (6-n) F_n=12$ where the sum runs over the n-polygon faces $F_n$. For polyhedra made entirely of hexagons and pentagons the number of pentagons must be 12 while the number of hexagons is not restricted by the theorem. The $C_{60}$ family is made by polyhedra with 12 pentagons and an increasing number of hexagons maintaining the symmetry of the icosahedron. The first fictitious gauge field that was proposed in the actual context appeared when trying to extract the electronic spectrum of the fullerenes from the low energy description of graphene based on the Dirac equation \cite{GGV92,GGV93A,GGV93}.

\section{Topological aspects.}
\label{sec_topo}

Here we will review some aspects of the topology the momentum space in fermionic systems on a lattice. We will use graphene and the Quantum Hall effect as test systems to see these ideas in action.

\subsection{Topology and Condensed Matter Physics.}

In a very broad sense, topology is the branch of mathematics devoted to the qualitative study of forms and their classification in terms of the invariance under certain transformations. For instance we can globally classify three dimensional surfaces by counting the number of holes they have. It is very well known that the surface of a sphere is topologically equivalent to the surface of  a bottle whereas  a donut is topologically equivalent to a mug or a pipe. In condensed matter physics this classification of objects in terms of shapes is useful although not so obvious as the examples mentioned before. During the last century, the classification of the quantum phases of matter has been done based on three major pillars: the band theory of solids, the Landau theory of the Fermi liquid, and the concept of spontaneous symmetry breaking.  Topology plays a role in the third pillar in terms of the notion of the order parameter and homotopy theory. It is not our intention to develop these ideas here and we refer to the literature \cite{M79} for a general and complete overview of the problem. We will nevertheless mention some of the most the prominent examples: The planar vector model whose isolated vortices in the vector order parameter cannot be continuously deformed to any smooth spin configuration in the plane.  The liquid helium where the vortices appear in the phase of the superconducting wavefunction. The theory of defects in crystals \cite{N02} and so on.

\subsection{Quantum Hall effect in graphene.}
\label{sec_hall}
As it was discussed in Sect. \ref{sec_features} the bandstructure of graphene possess two distinctive features when compared with other realizations of a two dimensional electron gas: the linear shape at low energies and the presence of a $SU(2)$ quantum number different from the spin, often referred in the literature as the valley spin or pseudo-spin. These two distinctions of graphene are the key features when we consider the topological properties of this system related to the Berry phase. 

Let us consider a single layer of graphene under the effect of an external homogeneous magnetic field, described by the vector potential $A=B(-y,0,0)$. In the basic description of the Quantum Hall Effect (QHE) in graphene we do not need to use the full description in terms of the two species (valleys) of Dirac fermions just because the external magnetic field breaks time reversal symmetry and both species will contribute equally to the Hall conductivity. We will extensively use of the so called magnetic length $l_{B}$ defined as $l_{B}=\frac{1}{\sqrt{eB}}$.

The low energy hamiltonian for one valley in graphene can be written as ($\hbar=1$)

\begin{equation}
H=v\bm{\sigma}\bm{\Pi}=v\left(\sigma_{x}\Pi_{x}+\sigma_{y}\Pi_{y}\right),\label{Diraceq}
\end{equation} 

with $\Pi_{j}=-i\partial_{j}+eA_{j}$. In terms of the ladder opperators $a=\frac{l_{B}}{\sqrt{2}}(\Pi_{x}-i\Pi_{y})$ and $a^{\dagger}=\frac{l_{B}}{\sqrt{2}}(\Pi_{x}+i\Pi_{y})$, satisfiying $[a,a^{\dagger}]=1$, the hamiltonian $H$ reads

\begin{eqnarray}
H=\frac{\sqrt{2}v}{l_{B}}\left(\begin{array}{cc}
0 & a\\
a^{\dagger} & 0
\end{array}\right)	\label{magnetichamiltonian}.	   
\end{eqnarray}

It is important to note that here we are considering the situation of a perfect crystalline layer of graphene with inversion symmetry so there is no term proportional to $\sigma_{z}$. The presence of such important term and similar ones will be considered later on. The Schr\"{o}dinger equation with the hamiltonian (\ref{magnetichamiltonian}) can be easily solved in terms of the solutions of the harmonic oscillator. The important difference here with the standard two dimensional electron gas is that the hamiltonian (\ref{magnetichamiltonian}) is a two dimensional matrix, and the eigenvectors are two dimensional spinors of the form $\psi_{n}=(u_{n},v_{n})^{T}$. In terms of the second component of this spinor, $v_{n}$, and the number operator $\hat{n}\equiv a^{\dagger}a$, we conclude that $\frac{2v^{2}}{l^{2}_{B}}\hat{n}v_{n}=E^{2}v_{n}=n v_{n}$, with $n$ being an integer. Without entering in more details, we see that for each value of $n$ we get two eigenvalues $E_{\pm,n}=\pm v \sqrt{2 |n|}/l_{B}\equiv \pm\omega_{c}\sqrt{|n|}$. An important observation from this equation is that there is an eigenstate corresponding to the value $n=0$ with no analogue in the standard  non relativistic two dimensional case \cite{Sch91}. 

In order to extract the information concerning the topology of this system we shall calculate the effective action through the use of the Landau levels structure \cite{H84}. We will consider the situation of zero temperature and finite chemical potential $\mu$. Within a path-integral approach, the effective action is calculated through $S_{eff}=-i \ln \mathcal{Z}$. After integrating out the fermion fields in $\mathcal{Z}$ the effective action reads

\begin{equation}
S_{eff}=\int\frac{d\omega dk}{4\pi^{2}}\sum^{\infty}_{n=-\infty}\ln \left(\omega-\mu-E_{n}\right).\label{effactionLandau}
\end{equation}

Because $E_{n}$ does not depend on the momentum $k$ each Landau level is (highly) degenerate. The degeneracy of each Landau level was first stimated by Landau himself to be $\frac{|eB|}{2\pi}$. By adding an small parameter $\eta\rightarrow 0^{+}sign(\omega)$ in the argument of the logarithm, the integration over the $\omega$ variable can be performed, leading to the following effective lagrangean

\begin{eqnarray}
\mathcal{L}_{eff}=\frac{|eB|}{4\pi}\sum^{+\infty}_{n=-\infty} |E_{n}-\mu|,\label{effectivelagrangean1}
\end{eqnarray}

or, splitting all the contributions from $n=0$ and $n\neq 0$ we get

\begin{eqnarray}
\mathcal{L}_{eff}=\frac{|eB|}{4\pi}|\mu|+\frac{|eB|}{4\pi}\sum^{+\infty}_{n=1} \left(|\mu+\omega_{c}\sqrt{|n|}|+|\mu-\omega_{c}\sqrt{|n|}|\right).\label{effectivelagrangean2}
\end{eqnarray}
The averaged particle number can be calculated from (\ref{effectivelagrangean2}) by taking the derivative of $\mathcal{L}_{eff}$ with respect of its conjugate variable $\mu$:
\begin{eqnarray}
n=\frac{\partial \mathcal{L}_{eff}}{\partial \mu}=&\frac{|eB|}{4\pi}\sum^{\infty}_{n=1}(sign(\mu+\omega_{c}\sqrt{|n|}) \\ \nonumber
&+ sign(\mu-\omega_{c}\sqrt{|n|})
+\frac{|eB|}{4\pi}sign(\mu).\label{averagedensity}
\end{eqnarray}
Before entering in the details of eq.(\ref{averagedensity}) let us see what is the actual meaning of the magnitude $n$ as a function of the magnetic field $B$. From (\ref{averagedensity}) we see that all the terms are proportional to $B$ (without loss of generality we shall assume that $B>0$): $n=\sigma_{H} B$, being $\sigma_{H}$ the Hall conductivity. We know that the average particle number is nothing but the temporal component of the gauge invariant electronic current: $n\equiv J^{0}$, and that the magnetic field $B$, being the $z-$component of the vector magnetic field $\mathbf{B}$, can be written in terms of the vector potential $A_{\mu}$, $B=\mathbf{B}_{z}=\partial_{2}A_{1}-\partial_{1}A_{2}$. Using these considerations and noticing that the system is gauge invariant, the previous relation $n=\sigma_{H} B$ can be written in a the following gauge invariant way:
\begin{equation}
J^{\mu}=\sigma_{H}\epsilon^{\mu\nu\rho}\partial_{\nu}A_{\rho}.\label{hallcurrent}
\end{equation}
 
The equation of motion (\ref{hallcurrent}) which describes the effective dynamics of the electrons under the effect of an external magnetic field can be derived from the well known Chern-Simons action:

\begin{equation}
\mathcal{S}_{CS}=\frac{\sigma_{H}}{4\pi}\int d^{3} x\epsilon^{\mu\nu\rho}A_{\mu}\partial_{\nu}A_{\rho}-J^{\mu}A_{\mu}.\label{CSlagrangean}
\end{equation}

\begin{figure}
\includegraphics[scale=0.5]{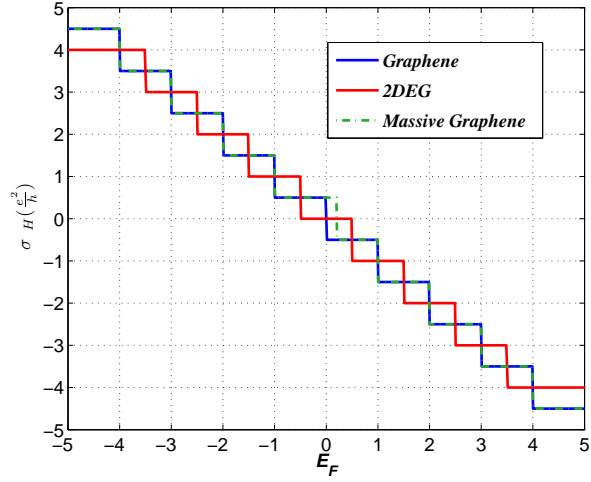}
\caption{Hall conductivities for graphene \emph{in the single cone approximation} (continuous red line) and the two dimensional electron gas (blue dashed line) as a function of the filling factor $\nu$. The (green) dotted dashed line corresponds to the case of graphene with a mass.}
\label{Hallsteps}
\end{figure}

So from the knowledge of the Landau levels, we have arrived to the well established effective lagrangean describing the physics of the integer QHE. The behaviour of the Hall conductivity $\sigma_{H}$ can be  obtained by inspection  of eq.(\ref{averagedensity}) and it is plotted in fig.(\ref{Hallsteps}). The first two terms in the right hand side of (\ref{averagedensity}) have the form of a staircase changing its value by integers when the chemical potential crosses the energy of each Landau level. This is exactly the behaviour found in the standard two dimensional electron gas, plotted also in fig(\ref{Hallsteps}). The difference is thus the last term, $sign(\mu)$, wich adds a remarkable change of one half to the entire Hall conductivity. Being remarkable, this is actually not in odds with the result of Thouless and others, who showed that $\sigma_{H}$  must take integer values. The solution of the paradox is that we actually have two Dirac points in the bandstructure of graphene (garanteed by the Nielsen-Ninomiya theorem) and we have to multiply the entire result by two Dirac species and by two spin orientations, so the actual values taken by the Hall conductivity are ($N\in \mathbf{Z}$)

\begin{equation}
\sigma_{H}=2\frac{e^{2}}{h}\left(2N+1\right).\label{Hallquantized}
\end{equation}

\subsection{The Callan-Harvey mechanism and the edge states in the QHE.}

We will comment on two aspects concerning the effective Chern Simons lagrangean. One quite general, related to the structure of the effective CS lagrangean (\ref{CSlagrangean}) and the existence of chiral states at the boundaries of the system, and the second, more specific to graphene, regarding the appearance of the parity anomaly when a mass term is added to the hamiltonian (\ref{Diraceq}).

It is easy to check that when dealing with ideally infinite systems, the laction defined in (\ref{CSlagrangean}) is gauge invariant under the gauge transformation $A_{\mu}\rightarrow A_{\mu}+\partial_{\mu}\Lambda$, as long as $\Lambda$ is some well behaved function in the entire space. However, when a bounded region is considered (or more generally, a domain wall), the action (\ref{CSlagrangean}) is no longer gauge invariant:

\begin{eqnarray}
S'_{CS}=S_{CS}-\frac{\sigma_{H}}{4\pi}\int_{\partial \Omega} d^{2}x \Lambda(x) \epsilon^{\mu\nu\rho}\hat{n}_{\mu}(x)\partial_{\nu}A_{\rho},\label{noninvariance}
\end{eqnarray}
 where $\hat{n}_{\mu}(x)$ stands for a unit vector perpendicular to the boundary at any point. Without loss of generality, we can choose the boundary to lie on the line $y=0$ and thus the vector $\hat{n}_{\mu}$ to point in the $(0,0,1)$ direction. Then, the second term in the right hand side of (\ref{noninvariance}) becomes ($i,j=0,1$)

\begin{equation}
\delta S=\frac{\sigma_{H}}{4\pi}\int d^{2}{x}\Lambda(x)\epsilon^{ij}\partial_{i}A_{j}=\frac{\sigma_{H}}{4\pi}\int d^{2}x\Lambda(x)\epsilon^{ij}F_{ij}.\label{noninvariance2}
\end{equation} 
 
For a generic regular function $\Lambda$ inside the bounded region $\Omega$, the right hand side of (\ref{noninvariance2}) is
nothing but the expression for the $(1+1)$ chiral anomaly, so through the noninvariance of the Chern Simons action in a region with boundaries, we reach the interesting conclusion that we can recover the gauge invariance for the system in a bounded region if we add to the system a chiral fermion whose anomalous contribution to the action exactly cancels the noninvariant piece of (\ref{noninvariance2}). So through this mechanism, known as the Callan-Harvey mechanism\cite{CH85}, we can rationalize the existence of chiral edge states at the boundaries of a system in the quantum Hall regime. Because $\sigma_{H}$ is an integer, we see that the value of the quanticed Hall conductance tell us the number of chiral edge states we have in our system. Let us stress that this argument lies only on the gauge noninvariance of the Chern Simons action and is applicable to any two dimensional fermionic system, graphene or the standard two dimensional electron gas. 

\subsubsection{Parity anomaly and QHE.}
\label{sec_berry}

In Quantum Electrodynamics in $2+1$ dimensions an interesting phenomenon emerges when a single species of massive Dirac fermions in the continuum is considered\cite{DJT82}. The action for such a system coupled to an abelian gauge field $A_{\mu}$ is (again, in units $h=c=1$)

\begin{equation}
S=\int d^{3}x\bar{\psi}\left(\gamma^{\mu}(i\partial_{\mu}+eA_{\mu})-m\right)\psi.\label{QEDaction}
\end{equation} 

It is widely well known that the mass term in the action (\ref{QEDaction}) breaks time reversal symmetry and parity just because in order to be symmetric under any of these two symmetries, $m$ must change to $-m$, fact that only happens for $m=0$. This lack of invariance under these two discrete symmetries will have as a consequence the emergence of a Chern Simons term in the one loop effective action for the field $A_{\mu}$. 

The one loop contribution to the polarization function is given by the bubble diagram, and it takes the form, in terms of the Feynman propagator for the Dirac field:

\begin{equation}
\Pi_{0}^{\mu\nu}(k)=e^{2}Tr \int \frac{d^{3}q}{8\pi^{3}}\gamma^{\mu}\frac{1}{\gamma^{\rho}(q_{\rho}+k_{\rho})-m}\gamma^{\nu}\frac{1}{\gamma^{\sigma}q_{\sigma}-m}.\label{polarization1}
\end{equation} 

Instead of evaluating (\ref{polarization1}) in its full glory, let us take a short cut to see  how the Chern Simons term emerges. We see from (\ref{CSlagrangean}) that the Chern Simons term is linear in derivatives, so in terms of momenta, it will be linear in $k$. So We can expand (\ref{polarization1}) in powes of $k$ and calculate the linear term. Differenciating (\ref{polarization1}) and setting $k=0$ it is found that the coefficient is given by the following integral

\begin{eqnarray}
e^{2}Tr\int\frac{d^{3}q}{8\pi^{3}}\gamma^{\mu}\frac{1}{\gamma^{\sigma}q_{\sigma}-m}\gamma^{\mu}\frac{1}{\gamma^{\sigma}q_{\sigma}-m}\gamma^{\rho}\frac{1}{\gamma^{\sigma}q_{\sigma}-m}. \label{polarization2}
\end{eqnarray}

After performing the trace, among all the terms in (\ref{polarization2}) there is one proportional to $m^{3}\epsilon^{\mu\nu\rho}$:

\begin{equation}
\epsilon^{\mu\nu\rho}\frac{e^{2}m^{3}}{8\pi^{3}}\int d^{3}p\frac{1}{(p^{2}-m^{2})^{3}}.
\end{equation}
This integral can be easily evaluated taking into account that it depends on $m^{2}$ and it is insensitive to the sign of the mass (also, $m^{3}=m^{2}·m=|m|^{2}·m$). The final result is that the effective action for the gauge field contains a Chern Simons term through the one loop contribution to the propagator:

\begin{equation}
\mathcal{L}=\frac{e^{2}}{4\pi}\frac{m}{|m|}\epsilon^{\mu\nu\rho}A_{\mu}k_{\nu}A_{\rho}.\label{CSQED3}
\end{equation}

We must stress that this expression has been calculated after expanding the polarization bubble to first order in $k$, so it is not completely \emph{exact}. It can be proven however that, first, the full dependence of this term with $k$ is on the form $f(k)\epsilon^{\mu\nu\rho}k_{\nu}$, being $f(k)$ a function only on the modulus of $k$, so the structure of the Chern Simons is maintained but with a more subtle dependence with the momentum, and second, it was proven by Coleman and Hill that the Chern Simons part of the polarization operator given by the one loop contribution (\ref{polarization1}) does not get renormalized by higher loop contributions, so the coefficient depending on $m$ in (\ref{CSQED3}) is exact in this model.

Let us set aside this result for a while and go back to the original problem of graphene in the single cone and low energy approximation under the effect of an external magnetic field, but this time adding to the original hamiltonian (\ref{Diraceq}) a mass term. We can play the same game as we did before, that is, calculate the new Landau level spectrum, project the effective action onto this Landau level basis for a finite chemical potential if necessary, calculate the average particle number as the derivative of the effective lagrangean with respect the chemical potential, and directly read the expression for the Hall conductivity. 

In this case, the Landau level energies are given by the following expression:

\begin{equation}
E_{n}=\pm \frac{v}{l_{B}}\sqrt{2|n|+m^{2}}, \label{LLspectrumfinitemass}
\end{equation}

but this time this relation holds for $n\neq0$ only. From the pair of solutions that naively would lead the value $n=0$ only one of them is actually normalizable (the sign of $B$ determines which solution is normalizable). This means that, as in the case of $m=0$, there are pairs of Landau levels related by $E_{-n}=-E_{n}$ except for the case $n=0$, which gives an unpaired state. In this case, the average number of particles $n$ is given by the rather cumbersome expression

\begin{eqnarray}
n=-\frac{eB}{4\pi}\frac{m}{|m|}\theta(|m|-\mu)
+\frac{|eB|}{2\pi}\left(I(\mu,m,B)+\frac{1}{2}\right)\theta(\mu-|m|),\label{massiveparticlenumber}
\end{eqnarray}

where $I$ is a integer valued function of the absolute value of the mass, the field, and the chemical potential. What is important here is the limit $\mu\rightarrow 0$ for the contribution coming from the unpaired Landau level for which the former expression reduces to 

\begin{equation}
n=-\frac{eB}{4\pi}\frac{m}{|m|}, 
\end{equation}

which of course, can be written in a gauge invariant phasion through a Chern Simons lagrangean with Hall conductivity $\sigma_{H}$:

\begin{equation}
\sigma_{H}=\frac{e^{2}}{4\pi}\frac{m}{|m|}.\label{Hallmassive}
\end{equation}

At this time it can be shown that the gauge invariant current is parity odd (recall the difference with the contribution of the zero mode in the massless case: there the magnetic field entered through its absolute value. This is another way to see that there is no parity anomaly in the massless limit, although the time reversal symmetry is already broken and a genuine Chern Simons term appears)\cite{Sch91,A85}.

Comparing the coefficient in front of (\ref{CSQED3}) and the value for the Hall conductivity for the lowest Landau level, one cannot avoid to wonder if there is any connection between the QHE for massive Dirac fermions and the $QED_{2+1}$ theory. 

The connection, if exists at all, must be quite deep, because despite of the superficial resemblance between the two Chern Simons theories, some few differences are apparent. In the case of the QHE, both in graphene and in the standard two dimensional electron gas, the coefficient in front of the Chern Simons action changes its value when the chemical potential eventually crosses a new Landau level, while in $QED_{2+1}$ the Hall conductivity is fixed. Also, in the case of the QHE, the expression for the  Chern Simons lagrangean is the exact one, while in $QED_{2+1}$, as we said, the momentum dependence in the expression (\ref{CSQED3}) is valid up to first order in $k$\cite{CoH85}. The third and definite difference is that in $QED_{2+1}$ the Chern Simons term is absent if we have pairs of Dirac species related by time reversal symmetry (or parity). We can see this in the case of graphene with a mass, where we do have two species of Dirac fermions related by time reversal symmetry, so in absence of any time reversal breaking perturbation, the two copies would lead to corresponding Chern Simons copies, but with opposite sign, letting a vanishing Hall conductivity. However, in the presence of the external magnetic field, both species contribute with the same coefficient to the Hall conductivity, so we have a non-zero $\sigma_{H}$.

%This situation is somehow analogous to the reaction of the conservative part of the Victorian english society in the nineteenth century about the discoveries of Charles Darwin. The message caught by the people (and the Church) from the Darwin's work was that the man descended from the apes, or the man and the ape were cousins, while what Charles Darwin actually said was that the human being and the apes shared common ancestors. Here the Chern Simons term in the QHE and the Chern Simons term in $QED_{3}$ are not even cousins, but they have a common root: a nontrivial topology coming from the Berry phase (we will see also that there are other kinds of topologically nontrivial systems which enjoy time reversal symmetry).
  
We will see that the common root between the systems with a Chern Simons term in their effective action is a nontrivial topology coming from the Berry phase. We will also see that in fact playing with the Berry phase\cite{B84}, we can find time reversal symmetric systems with a topological term in their effective actions similar to some extent to the Chern Simons term.

\subsection{The Berry phase and the Integer QHE.}

We have seen in the previous section how we can describe the low energy effective action governing the physics of a system (both graphene and the standard two dimensional electron gas) under the effect of an external homogeneous magnetic field by means of a Chern Simons term in the lagrangean. We obtained this term under the assumption that the system was in the quantum Hall regime, that is, when the magnetic field was so strong to induce the presence of Landau levels. The quantization of the Hall conductivity seemed to follow from the existence of such quantized levels. Although this picture is true, the presence of Landau levels is not the ultimate reason for the quantized nature of the Hall conductivity (although extremely hard to find in condensed matter physics, the example of a single species of planar Dirac fermions is the proof of that). Also, in absence of interactions, it is hard to find crystalline solids where the time inversion symmetry is absent so we cannot expect to find any term similar to (\ref{CSQED3}) in any way by looking for situations where the one particle spectrum is described by Landau levels, that is, having flat bands in the spectrum does not necessarily mean that those bands will display a Chern Simons-like term in the effective lagrangean. 

In order to understand the ultimate reason of the quantization of the Hall conductivity, and envisage any possibility of finding similar physics in condensed matter systems, we have to go to the lattice and analyze the same problem, but this time not relying on a continuum approximation. We will see that the concept of Berry phase is behind of this quantization, that it has a topological nature, and it is the element that eventually will allow us to extend these concepts to time reversal invariant systems.

\subsubsection{Berry phase and the quantization of the monopole charge.}

Let us begin this section with a brief review of the concept of Berry phase. We will follow a rather pedagogical way and introduce the concept of geometrical phase following the original line of thinking from Berry's original work\cite{B84}. Berry originally considered the effect on the phase factor of the wavefunction when an slow change was performed in the parameters of the problem.

Consider a quantum one particle system described by the hamiltonian $H(\mathbf{s})$ depending on the set of parameters $\mathbf{s}=(s_{1},..,s_{n})$. We will assume that the hamiltonian is time dependent through the parameters $\mathbf{s}(t)$. If the parameters $\mathbf{s}$ change slowly with time, the Schroedinger equation can be solved in the adiabatic limit. If the system of eigenstates in principle is not degenerate, at any given $t$ we have

\begin{equation}
|n(t)\rangle=e^{-i\int^{t}_{0}d\tau E_{n}(\tau)+i\gamma_{n}(t)}|n(0)\rangle.\label{timeevolvingstate}
\end{equation} 

The first term in the exponential is the standard dynamic phase while the second term is a yet undertermined phase allowed by the adiabatic theorem. If (\ref{timeevolvingstate}) is a solution of the Schoedinger equation $H(\mathbf{s}(t))|n(t)\rangle=E_{n}(t)|n(t)\rangle$, then the phase factor $\gamma_{n}$ adquires the form

\begin{equation}
\gamma_{n}(t)=i\int^{t}_{0}d\tau \langle n(\tau)|\frac{\partial}{\partial\tau}|n(\tau)\rangle,\label{Berryphase1}
\end{equation}
or in terms of the parameters $\mathbf{s}$,
\begin{equation}
\gamma_{n}(t)=i\int^{\mathbf{s}(t)}_{\mathbf{s}(0)}d\mathbf{s}\langle n(\mathbf{s})|\nabla_{\mathbf{s}}|n(\mathbf{s})\rangle=i\int^{\mathbf{s}(t)}_{\mathbf{s}(0)}d\mathbf{s}\mathcal{A}_{n},\label{Berryconnection}
\end{equation}

with the obvious definition $\mathcal{A}_{n}=\langle n|\nabla_{\mathbf{s}}|n\rangle$. The vector $\mathcal{A}_{n}$ is usually termed the Berry connection.

The interesting situation comes when we consider the particular case of a system with a multidimensional parameter set $\mathbf{s}$ and a time evolution where at time $t=T$, some loop $\Gamma$ in the parameter space is performed and therefore $\mathbf{s}(T)=\mathbf{s}(0)$. In this situation, the Berry phase $\gamma_{n}$ reads (after using the Stokes theorem)

\begin{equation}
\gamma_{n}=i\oint_{\Gamma} d\mathbf{s}\mathcal{A}_{n}=i\int_{S}d\mathbf{S}\nabla_{\mathbf{s}}\times\mathcal{A}_{n}=i\int_{S}d\mathbf{S}\cdot\mathcal{B}_{n}.\label{Berrycurvature}
\end{equation}

In the absence of any singularity in the parameter space or in the vector field $\mathcal{A}_{n}(\mathbf{s})$, any loop $\Gamma$ can be continuously deformed to zero and there is no Berry phase effect in the wavefunction. The funny situation comes when the parameter space is itself nontrivial or the field $\mathcal{A}_{n}$ shows some singularities. Let us recall that although we have made use of the adiabatic hypothesis, it is actually not needed since the time $t$ (or $\tau$)  disappears from the expressions (\ref{Berryphase1}), (\ref{Berryconnection}), and (\ref{Berrycurvature}) and all the expressions are written in terms of geometrical quantities defined in the parameter space.

We can follow the original paper of Berry and transform (\ref{Berrycurvature}) a little bit more:

\begin{eqnarray}
\gamma_{n}=i\int d\mathbf{S}\nabla_{\mathbf{s}}\times\langle n|\nabla_{\mathbf{s}}n\rangle
&=i\int d\mathbf{S}\langle\nabla_{\mathbf{s}}n|\times|\nabla_{\mathbf{s}}n\rangle \\ \nonumber
&=i\int d\mathbf{S}\sum_{n\neq m}\langle\nabla_{\mathbf{s}}n|m\rangle\times\langle m|\nabla_{\mathbf{s}}n\rangle.\label{Berryphase2}
\end{eqnarray}
If we consider the situation where the eigenvalues $E_{n}$ are not degenerate, for $n\neq m$, we have $(E_{n}-E_{m})\langle m|\nabla_{\mathbf{s}}|n\rangle=\langle m|\nabla_{\mathbf{s}}H|n\rangle$ we find the important final relation:
\begin{eqnarray}
\gamma_{n}=-i\int_{S}d\mathbf{S}\sum_{n\neq m}\frac{\langle n|\nabla_{\mathbf{s}}H|m\rangle\times\langle m|\nabla_{\mathbf{s}}H|n\rangle}{(E_{n}-E_{m})^{2}}.\label{intBerryphase}
\end{eqnarray}

Although in the expression (\ref{intBerryphase}) we have written the Berry phase $\gamma_{n}$ making use of the non degeneracy of the eigenstates of the system, the previous expressions can be used to calculate this phase in the situations when degeneracies in the parameter space exist. We will use the low energy hamiltonian in graphene to illustrate this situation and as a toy model to show that the Dirac quantization of the monopole charge is actually a topological quantization due to the existence of a nontrivial Berry phase in that system.

This time we will take into account both species of massless Dirac fermions in the low energy sector of the spectrum in graphene (our parameter space here will be the momentum space).  As usual, the effective hamiltonian around the Fermi points $K,K'$ can be written as

\begin{equation}
H_{K,K'}=v\left(\tau_{z}\sigma_{x}k_{x}+1_{\tau}\sigma_{y}k_{y}\right).\label{effDirac}
\end{equation}

The eigenstates around both Fermi points are $E(\mathbf{k})=s v|\mathbf{k}|$, and the eigenstates are collectively written as

\begin{eqnarray}
\Psi_{s,\tau}=\frac{1}{\sqrt{2}}
\begin{pmatrix} 1\\s\tau e^{i\tau\theta_{\mathbf{k}}}\\
\end{pmatrix},\label{eigenstates}
\end{eqnarray}

where $s=\pm 1$ and $\tau=\pm 1$ stand for band index and Fermi points, respectively, and $\theta_{\mathbf{k}}$ is the angle defined by the wave vector $\mathbf{k}$ with the horizontal axis. The Berry connection (\ref{Berryconnection}) is easily calculated to be ($i,j=1,2$)

\begin{equation}
\mathcal{A}^{i}_{s,\tau}(\mathbf{k})=\frac{\tau}{2}\frac{\epsilon^{ij}k_{i}}{|\mathbf{k}|^{2}}.\label{monopolegraphene1}
\end{equation}
In terms of the Berry curvature, we get
\begin{equation}
\mathcal{B}^{z}_{s,\tau}=\frac{\tau}{2}\delta^{(2)}\left(\mathbf{k}\right),\label{monopolegraphene2}
\end{equation}
that is, both species carry a quantized monopole charge of value one half (sitting at the $K$ and $K'$ points), and it has opposite value for each specie. In this particular case, the total magnetic charge is the sum of the two charges in (\ref{monopolegraphene2}) and this zero, being the Nielsen-Ninomiya theorem behind this particular result\cite{NN81}. We have to make a comment here concerning the particular value the monopole charge takes. It is a general statement that around degenerate points the monopole charge takes half integer values. We will see in the next sections that in general this topological charge takes integer values instead. There is no contradiction since, as we have said, the Nielsen Ninomiya theorem ensures to have an even number of degenerate points, and the total charge will be zero, if the degeneracy points are related by parity or time reversal invariance, or some integer, because the sum of even number of half integer charges is an integer, if the points are not related by theses symmetries. It is common to find these degeneracies lifted by some allowed perturbation. When the total charge is zero or an integer depends on the nature of this perturbation.  

This case exemplifies how a nontrivial structure in the wavefunctions leads to a vector field (Berry curvature) which is singular at some points of the parameter space, and the total integral of the curvature associated to this singular connection over a closed surface (in this case, any two disjoint spheres containing separately the points $K$ and $K'$) leads to quantized value.

\subsubsection{QHE on the lattice and the magnetic Brillouin zone.}

Let us consider, for this time, fermions defined on the lattice whose dynamics under the effect of a magnetic field are described by the following tight binding hamiltonian with coupling to nearest neighbours. In what concerns the basics, there is no much more complication if we study the problem in the square lattice or in the honeycomb lattice instead\cite{BHZ07}. We will follow the original work of Hofstadter\cite{H76}:

\begin{equation}
H=\sum_{<i,j>} t(\mathbf{R}_{i},\mathbf{R}_{j})C^{+}(\mathbf{R}_{i})C(\mathbf{R}_{j}),\label{TBham}
\end{equation}
where $\mathbf{R}_{i}$ labels the sites of the square lattice, and the four nearest neighbours are $\mathbf{R}_{j}=\left\lbrace\mathbf{R}_{i}\pm a\mathbf{e}_{x},\mathbf{R}_{i}\pm a\mathbf{e}_{y}\right\rbrace$. The hopping term $t(\mathbf{R}_{i},\mathbf{R}_{j})$ is defined by the standard Peierls substitution:

\begin{eqnarray}
t(\mathbf{R}_{i},\mathbf{R}_{j})=te^{ie\int^{\mathbf{R}_{j}}_{\mathbf{R}_{i}}\mathbf{A}(\mathbf{r})d\mathbf{r}}.\label{peierls}
\end{eqnarray}

In the Landau gauge $\mathbf{A}=B(0,x,0)$, the phase in (\ref{peierls}) is just $eBax$, where $x=a·m$ labels the $x-$component of the position $\mathbf{R}_{m}$ and $a$ is the lattice spacing. Due to translational invariance along the $y$ direction, we can rewrite the hamiltonian $H$ now reads

\begin{eqnarray}
H=&t\sum_{m}\sum_{k_{y}}C^{+}_{m}(k_{y})C_{m+1}(k_{y})+C^{+}_{m}(k_{y})C_{m-1}(k_{y})+ \\ \nonumber
&2\cos\left(2\pi\phi m+k_{y}a\right)C^{+}_{m}(k_{y})C_{m}(k_{y}),\label{HarperH}
\end{eqnarray}
where $\phi$ is the magnetic flux within each plaquette. When this magnetic flux per plaquette is a rational number, $\phi=\frac{p}{q}$,
the cosine term in (\ref{HarperH}) is invariant under the change $m\rightarrow m+q$ and so does the hamiltonian. The Bloch theorem thus tell us that $C_{m+q}=e^{ik_{x}qa}C_{m}$ for $q$ fixed by the magnetic flux. Our Schroedinger equation is a $q\times q$ matrix problem, with $q$ eigenvalues $E=E_{j}(\mathbf{k})$ with $j=1,...,q$. 

We can fix the periodicity of $E_{j}(\mathbf{k})$ by noting that the points $k_{x}$ and $k_{x}+\frac{2\pi}{aq}$ are equivalent in (\ref{HarperH}) so $0<k_{x}<\frac{2\pi}{aq},0<k_{y}<\frac{2\pi}{a}$, so we can define a Brillouin zone, which has the form of a torus (that the Brillouin zone has is a torus and thus a closed surface is one of the key ingredients for the quantization of the Hall conductivity, as we will see). The important message here is that although the presence of an external magnetic field prevents the system to enjoy the translational symmetry of the original system at $B=0$, we still have an enlarged translational symmetry, which is the same as to say that the original Brillouin zone is split up in $q$ copies. Also we have enlarged the number of bands and the new eigenstates are eigenvectors of $q$ components.

Instead of studying the consequences of this effective folding by the general properties of the solutions of (\ref{HarperH}), for our purposes it is enough to analyse two specific simple examples for two values of the pair $(p,q)$. 

Let us start for the pair $(p,q)=(1,2)$. In this case, the Schoredinger equation is a $2\times2$ matrix equation, which takes the particularly simple form of

\begin{eqnarray}
H=2t\left(\begin{array}{cc}
-\cos(ak_{y}) & \cos(ak_{x})\\
\cos(ak_{x}) &\cos(ak_{y})
\end{array}\right)	\label{pifluxham}.	
\end{eqnarray}

It is worth to mention that the situation with half of the flux per plaquette is known as the ''pi-flux phase". It is also worth to mention that for this particular choice of the flux the system is actually time reversal invariant, because as we said before, there is a periodicity in $\phi$ and the system with $\phi=\frac{1}{2}$ is equivalent to the system with $\phi=-\frac{1}{2}$. Nevertheless, the example $q=2$ is interesting because the spectrum takes the form $E_{\pm}(\mathbf{k})=\pm 2t\sqrt{\cos(ak_{x})^{2}+\cos({ak_{y}})^{2}}$.
 
\begin{figure}
\includegraphics[scale=0.5]{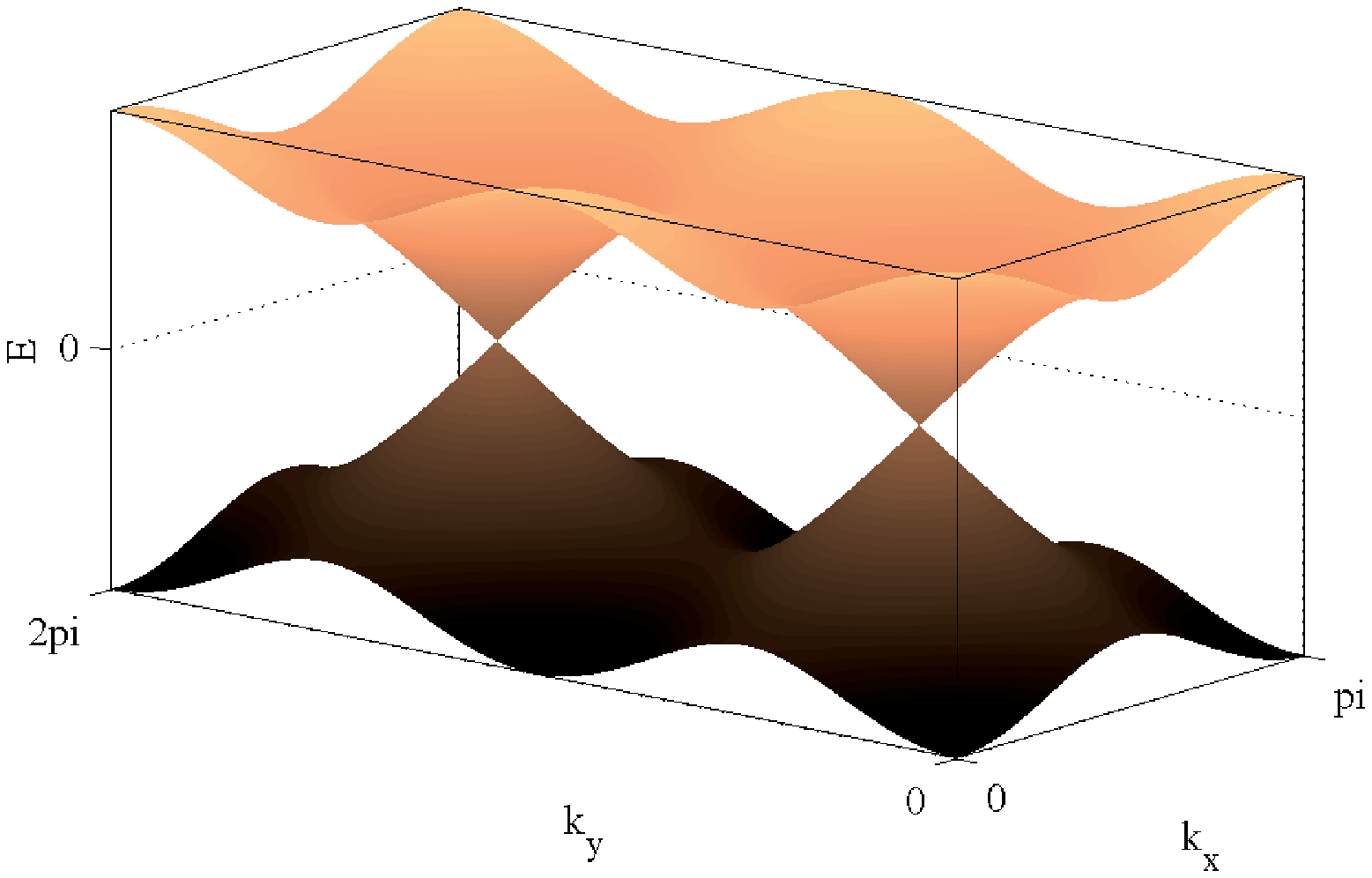}
\includegraphics[scale=0.6]{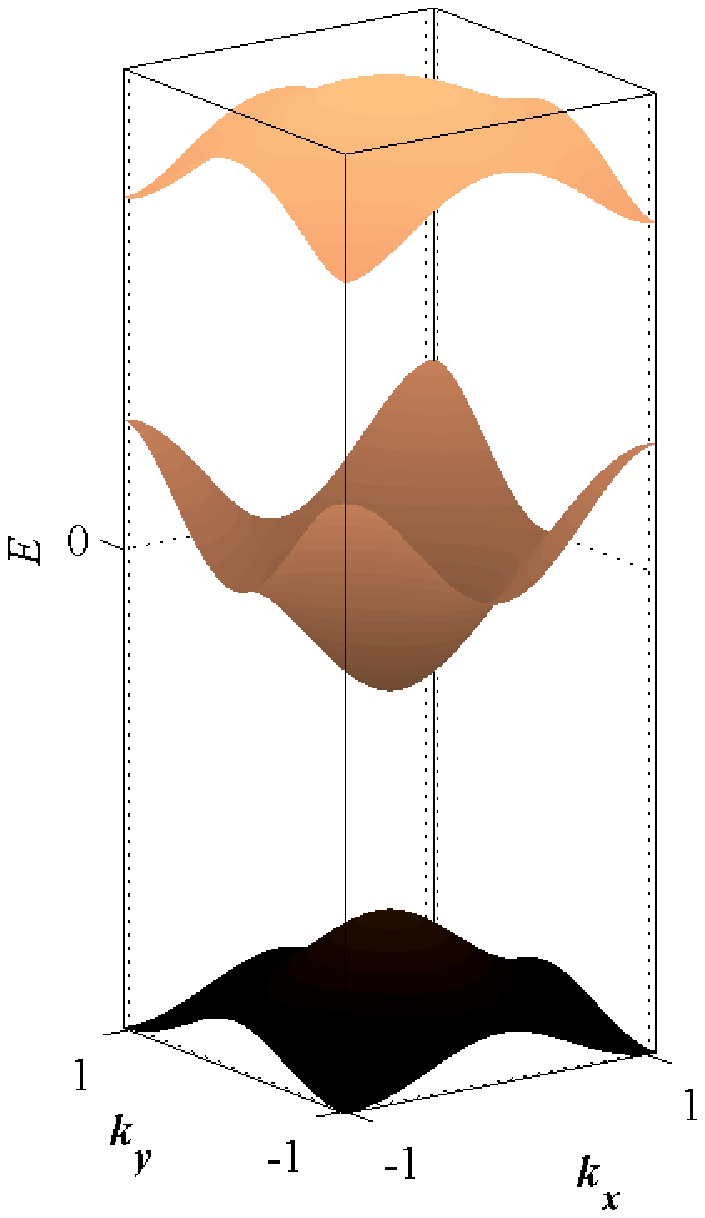}
\caption{Bandstructure of the pi-flux phase corresponding to the value $q=2$ (left) and
$q=3$ (right).}
\label{piflux}
\end{figure}
The spectrum consists in two bands that are degenerate at the points $(\pi/2,\pm\pi/2)$. Around these two points, the hamiltonian (\ref{pifluxham}) takes the familiar form:

\begin{equation}
H=v\tau_{z}\sigma_{x}\delta k_{x}+v\sigma_{z}\delta k_{y},\label{DiracHam}
\end{equation}
that is, the low energy hamiltonian of the system consists in two species of massless Dirac fermions, which are, interestingly, related by parity. This example illustrates a general property of (\ref{HarperH}) for even values of $p$: the band structure consists in gapped bands except for the two central bands that are degenerate at $q$ isolate points in the Brillouin zone (as in graphene, $\tau_{z}$ labels the two Dirac species). Around those points, the low energy system can be described by $q$ species of massless Dirac points\cite{WZ89}.  For odd values of $q$ the bands are globally separated by gaps in the entire Brillouin zone. In the case for $(p,q)=(1,3)$ the hamiltonian is the following $3\times3$ matrix

\begin{eqnarray}
H=t\left(\begin{array}{ccc}
\cos(ak_{y}+2\pi/3) & 1 & e^{3ik_{x}a}\\
1 & \cos(ak_{y}+4\pi/3) & 1\\
e^{-3ik_{x}a} & 1  & \cos(ak_{y}+6\pi/3)
\end{array}\right)	
\end{eqnarray}

The eigenvalues $E(\mathbf{k})$ are the solutions of the secular equation $$E^{3}-6t^{2}E=2t^{3}(\cos(3k_{x})+\cos(3k_{y}))$$ and they are depicted in fig.~\ref{piflux} (right) \cite{WZ89,ChN95}.
%\begin{figure}
%\includegraphics[scale=0.6]{bandsq3.eps}
%\caption{Bandstructure for the value $q=3$.}
%\label{bandsq3}
%\end{figure}

These two examples illustrate to what happens when electrons living on a lattice experience the effects of an external homogeneous magnetic field. The original Brillouin zone is split up in $q$ copies of a new magnetic Brillouin zone, which is still a torus. Also, the new wavefunctions become multicomponent Bloch functions $|u_{\mathbf{k}}\rangle$. This fact seems to be rather paradoxical if we again compare with the problem of Landau levels. In the previous sections we found that the spectrum of electrons in two dimensions under the effect of a external magnetic field consisted in discrete Landau levels, and, due to the explicit dependence of the vector potential $\mathbf{A}$ with the spatial coordinates, the invariance under translations was lost. A deeper look on the problem will tell us that this is not the case. Effectively, the physical momenta $\mathbf{\Pi}$ satisfy $[\Pi_{x},\Pi_{y}]=-ieB$, so two elements of the translation group do not commute in general. If $T(\mathbf{R})=e^{i\mathbf{\Pi}\mathbf{R}}$, then\cite{H76}

\begin{equation}
T(\mathbf{R}_{1})T(\mathbf{R}_{2})=T(\mathbf{R}_{2})T(\mathbf{R}_{1})e^{ie\mathbf{B}\cdot\left(\mathbf{R}_{1}\times\mathbf{R}_{2}\right)}.\label{magtrans}
\end{equation}
They do commute, however, when the phase $\Phi=e\mathbf{B}\cdot\left(\mathbf{R}_{1}\times\mathbf{R}_{2}\right)$ is an integer number, that is, when the flux passing through the area defined by $\mathbf{R}_{1}\times\mathbf{R}_{2}$ is an integer. In terms of the original primitive lattice vectors, this condition means that if the vectors are written in the original lattice basis, $\mathbf{R}_{1}=p\mathbf{a}$ and $\mathbf{R}_{2}=q\mathbf{b}$, we have the condition for flux per unit cell to be a rational number, $pq\phi_{plaquette}=n$, being $n$ some integer. This condition is precisely the one that we found at the beginning of this section, when we considered the eigenstates of (\ref{HarperH}). We then find that even when the original translational invariance is lost, we can still label the eigenstates within each Landau level in terms of a wave vector $\mathbf{k}$ when the magnetic flux per unit cell is a rational number. 

\subsubsection{Berry phase, Chern number, and the Hall conductivity.}

In this section we will close the circle and see the explicit connection between the Hall conductivity and the Berry phase constructed through the Bloch states corresponding to the new magnetic subbands $E_{j}(\mathbf{k})$. Our task is to find an expression that relates the Hall conductivity $\sigma_{H}$ and the Berry phase $\gamma_{n}$, and second, we need to demonstrate that the Berry connection defined by the Bloch eigenstates of (\ref{HarperH}) is a singular vector field at some points of the Brillouin zone, and thus the total monopole charge enclosed by this surface is an integer.

Let us employ linear response theory to find the desired relation between $\sigma_{H}$ and $\gamma_{n}$\cite{TKN82}. Assume that, together with the external homogeneous magnetic field $\mathbf{B}$ applied perpendicular to the system, we apply a weak electric field $\mathbf{E}$ pointing, say, in the $x$ direction. The corresponding vector field thus takes the form $A_{0}=E_{x} x$ and $\mathbf{A}=(0,Bx,0)$. We shall assume for simplicity that the magnetic subbands are globally separated by a gap in the entire Brillouin zone, so $E_{i}(\mathbf{k})\neq E_{j}(\mathbf{k})$ for every momentum $\mathbf{k}$ and $i\neq j$. Using the magnetic Bloch functions in the real space representation $|\psi_{n}(\mathbf{r})\rangle=e^{i\mathbf{kr}}|u_{n}(\mathbf{k})\rangle$ as the basis of unperturbed wavefunctions, simple perturbation theory gives the first correction for the perturbed wavefunctions:

\begin{equation}
|u_{n}(\mathbf{k})\rangle^{(1)}=|u_{n}(\mathbf{k})\rangle + i e E_{x}\sum_{n\neq m}\frac{|u_{m}(\mathbf{k})\rangle\langle u_{m}(\mathbf{k})|\frac{\partial}{\partial_{k_{x}}}|u_{n}(\mathbf{k})\rangle}{E_{n}(\mathbf{k})-E_{m}(\mathbf{k})}.\label{firstorder}
\end{equation}

In one particle band hamiltonians the velocity operator is defined as the gradient of the Hamiltonian, $\mathbf{v}(\mathbf{k})=\nabla_{\mathbf{k}}H(\mathbf{k})$ and the averaged electronic current is just the averaged velocity times the electric charge: $\langle\mathbf{j}\rangle=e\langle\mathbf{v}\rangle$. We then can use the expression for $\mathbf{v}(\mathbf{k})$ and the perturbed eigenstates (\ref{firstorder}) to calculate the first order correction to the $y$ component of the electronic current of the state $n$ (as long as $\langle \mathbf{v}_{n}\rangle$ is a real function, we must include also the term given by the complex conjugate of (\ref{firstorder})):

\begin{eqnarray}
\langle v_{n,y}\rangle^{(1)}=\langle u_{n}|v_{y}|u_{n}\rangle+i e E_{x} \int \frac{d^{2}\mathbf{k}}{4\pi^{2}}\sum_{n\neq m}\frac{\langle u_{n}|v_{y}|u_{m}\rangle\langle u_{m}|\partial_{k_{y}}u_{n}\rangle}{E_{n}-E_{m}}+c.c.\label{velocity1}
\end{eqnarray}

Again, we can make use here of the identity $(E_{n}-E_{m})\langle m|\nabla_{\mathbf{k}}|n\rangle=\langle m|\nabla_{\mathbf{k}}H|n\rangle$ as we did two sections ago and plug it in (\ref{velocity1}) giving

\begin{eqnarray}
\langle v_{n,y}\rangle^{(1)}=\langle u_{n}|v_{y}|u_{n}\rangle+i e E_{x} \int \frac{d^{2}\mathbf{k}}{4\pi^{2}}\sum_{n\neq m}\langle u_{m}|\partial_{k_{x}}u_{n}\rangle\langle u_{m}|\partial_{k_{y}}u_{n}\rangle-\langle u_{m}|\partial_{k_{y}}u_{n}\rangle\langle u_{m}|\partial_{k_{x}}u_{n}\rangle.\label{velocity2}
\end{eqnarray}

We readily read in (\ref{velocity2}) a term similar to the one in the last part of the expression (\ref{Berryphase2}) so the second term of (\ref{velocity2}) can be written in terms of a Berry connection\cite{B84}

\begin{equation}
\langle v_{n,y}\rangle^{(1)}=\langle u_{n}|v_{y}|u_{n}\rangle+i \frac{e E_{x}}{4\pi^{2}}\sum_{n} \int d^{2}\mathbf{k}\nabla_{\mathbf{k}}\times\mathcal{A}_{n}(\mathbf{k}).\label{velocity3}
\end{equation}

If the Fermi level lies on a gap above the band $E_{n}(\mathbf{k})$, the first term of the right hand side of  (\ref{velocity3}) do not contribute to the total current. Because we have the relation $\langle j^{i}\rangle=e\langle v^{i}\rangle$ and $\langle j^{i}\rangle=\sigma_{ij}E^{j}$, we read from (\ref{velocity3}) the value of the Hall conductivity:

\begin{equation}
\sigma_{H}\equiv \sigma_{xy}=\frac{e^{2}}{4\pi^{2}}\sum_{n}\int d^{2}\mathbf{k} i\nabla_{\mathbf{k}}\times\mathcal{A}_{n}(\mathbf{k})=\frac{e^{2}}{4\pi^2}\oint_{\Gamma}\mathcal{A}_{n}(\mathbf{k}),\label{Hallvalue}
\end{equation}

where $\Gamma$ is the loop defined by the corners of the magnetic Brillouin zone. This is the desired relation to prove: the Hall contribution to the conductivity is the total flux passing through the Brillouin zone of a Berry curvature. When we discussed the quantization of the charge of a magnetic monopole in the context of graphene, we found that if the Berry connection was singular at some points inside some closed surface, the monopole charge was quantized. Here we must show that the Berry connection in (\ref{Hallvalue}) defined through the magnetic Bloch eigenstates $|\psi_{n}(\mathbf{r})\rangle$ is singular at some point within the Brillouin zone.

The proof that there are points on the magnetic Brillouin zone were the Berry connection is singular is rather involved and we will only sketch it, referring the original work for details\cite{K85,F07}. 

We have stressed in the previous section that the magnetic Bloch wavefunctions that are solutions of (\ref{HarperH}) are no longer eigenstates of the translations of the original lattice, but they are of the new enlarged unit cell, that contains an integer value for the magnetic flux passing through the new magnetic plaquette. If we consider a loop consisting in such of one of these magnetic plaquettes, the phase of the magnetic Bloch wavefunction $|\psi_{\mathbf{k}}(\mathbf{r})\rangle\equiv \rho_{\mathbf{k}}(\mathbf{r})e^{i\theta_{\mathbf{k}}(\mathbf{r})}$ will have changed $2\pi p$: $\int d\mathbf{l}\nabla_{\mathbf{l}}\theta_{\mathbf{k}}(\mathbf{r})=2\pi p$ . This is nothing but the Aharonov-Bohm effect of an state encircling a region of the space where there is a nonzero magnetic flux. It means that the wavefunction has the form of a vortex at some point of the magnetic unit cell in order to accommodate the nonzero phase in a closed loop. As long as the wavefunction is not singular on its arguments, the wavefunction must be zero at the point where the vortex is, and for some point in the magnetic Brillouin zone, as long as the phase $\theta_{\mathbf{k}}(\mathbf{r})$ is a function of $\mathbf{k}$. Because the Bloch wavefunction has zeros on the torus defined by the magnetic Brillouin zone, the Berry phase will have points where it is ill-defined.

 Let us now assume that we can divide the magnetic Brillouin zone in two regions $R1$ and $R2$ separated by any loop $\gamma$, one containing the zero of the wavefunction. In the region that no contains the zero, the phase of the wavefunction is well defined globally, but in the region that contains the zero the phase might pick up any arbitrary value. Once chosen, we can allways choose the phase continuously in a small neigborhood around the zero with the condition that the phase cannot be defined globally. When we compare the wavefunctions away from the zero, the two choices of phase must be consistent through a gauge change, that is $\Psi_{I}=e^{if(\mathbf{k})}\Psi_{II}$, or, equivalently
 
\begin{equation}
\mathcal{A}_{n,I}(\mathbf{k})=\mathcal{A}_{n,II}(\mathbf{k})+\nabla_{\mathbf{k}}f(\mathbf{k}).\label{tophall1}
\end{equation}

By using the Stokes theorem, we can apply the definition of Hall conductivity (\ref{Hallvalue}) and write each contribution for the two regions:

\begin{equation}
\sigma_{H}=\frac{e^{2}}{4\pi^{2}}\int_{R1}d^{2}\mathbf{k}\nabla_{\mathbf{k}}\times\mathcal{A}_{n,I}+\frac{e^{2}}{4\pi^{2}}\int_{R2}d^{2}\mathbf{k}\nabla_{\mathbf{k}}\times\mathcal{A}_{n,II},\label{tophall2}
\end{equation}

and then, because $R1$ and $R2$ have opposite orientation,

\begin{equation}
\sigma_{H}=\frac{e^{2}}{4\pi^{2}}\int_{\gamma}(\mathcal{A}_{n,I}-\mathcal{A}_{n,II})d\mathbf{k}=\frac{e^{2}}{4\pi^{2}}\int_{\gamma}\nabla_{\mathbf{k}}f(\mathbf{k})d\mathbf{k}.\label{tophall3}
\end{equation}

The expression (\ref{tophall3}) means that the Hall conductivity counts the number of times the function $f(\mathbf{k})$ winds around $2\pi$ as $\mathbf{k}$ travels along the loop $\gamma$. Since this winding number does not change when the loop $\gamma$ is deformed (and also it contains the zero of the wavefunction), this winding number is a topological invariant, known as the first Chern number, and it takes integer values, which is the result we wanted to prove.
We have written the Hall conductivity as the surface integral of a Berry curvature, which is nothing but the total Berry phase calculated along a loop corresponding to the "boundary" of the Brillouin zone. In order to have a nonzero $\sigma_{H}$, the wavefunction $|\psi_{n}(\mathbf{k})\rangle$ must pick up a nonzero Berry phase when circulating along the mentioned loop. If the wavefunctions have no Berry phase, or equivalently, the Stokes theorem applied to this situation tell us that if $\mathcal{A}_{n}$ is not singular and we perform the integration over a surface without boundaries, the integral is zero. This is what happens with most of the band insulators. However, there is an intermediate case, which is when the system has an even number of singular points in the Berry connection, by requirements of time reversal symmetry, and also the winding number of those points have opposite value (we saw this in the case of the low energy hamiltonian of graphene). The total Berry phase is zero (and so does $\sigma_{H}$) but we are still able to find a modified topological quantity which takes into account the value of each singular point separately. We will analyse  this situation in the next section.

Another interesting observation concerning (\ref{Hallvalue}) is that when in normal band insulators the Fermi level lies on the gap, the Bloch wavefunction of the groundstate is made of linear combination of (maximally) localized Wannier functions, that is, the wavefunction of a state at the point $\mathbf{r}_{i}$ decays as (or faster than) $|\mathbf{r}-\mathbf{r}_{i}|^{-2}$\cite{T81}. It means that the Bloch wavefunction $|\psi(\mathbf{k})\rangle=e^{i\mathbf{kr}}|u(\mathbf{k})\rangle$ in the momentum space is an analytic function of $\mathbf{k}$. A Hall nonvanishing Hall conductivity means that we cannot define globally the phase of the Bloch wavefunction, so it will not be analytic \cite{BPetal07}. It thus implies that the ground state of a Hall system cannot be described in terms of maximally localized Wannier functions. This observation has interesting consequences in the modern geometrical theory of the electric polarization based on the Berry phase of maximally localized Wannier functions\cite{VK93}.

\subsection{Topological insulating phases in graphene.}
\label{sec_topomono}
In the previous sections we have found a relation between the Hall conductivity $\sigma_{H}$ and the Berry phase for the Quantum Hall effect. We have also seen that despite of the presence of an homogeneous magnetic field, we can still define a Hamiltonian which enjoys translational invariance (this translational invariance is not the same as the original system) and with the magnetic Bloch wavefunctions we can define a Berry connection, whose circulation around the perimeter of the Brillouin zone gives the quantized value of $\sigma_{H}$. However, at the beginning of the previous section, we compared the QHE with a single specie of massive Dirac fermion in two spatial dimensions finding in the two cases a Chern Simons term in the effective electromagnetic field theory, being the coefficient of the CS term the Hall conductivity.
Let us see how we get the same conclusion for $QED_{(2+1)}$ by using the Berry phase (\ref{intBerryphase}). We will make extensive use of this calculation in the subsequent sections.

The hamiltonian corresponding to the action (\ref{QEDaction}) takes the following form, in the momentum space ($v$ is the Fermi velocity and $\bm\sigma=(\sigma_{x},\sigma_{y})$): 

\begin{equation}
H=v\bm\sigma\mathbf{k}+m\sigma_{z}.\label{QEDham}
\end{equation}

The eigenvalues of this hamiltonian are, of course $E_{+}=-E_{-}=\sqrt{v^2|\mathbf{k}|^{2}+m^2}\equiv E_{\mathbf{k}}$, with the corresponding eigenvectors $|+\rangle=\sqrt{\frac{m+E_{\mathbf{k}}}{2E_{\mathbf{k}}}}\left(1,\frac{vk}{m+E_{\mathbf{k}}}\right)^{T}$ and $|-\rangle=\sqrt{\frac{m+E_{\mathbf{k}}}{2E_{\mathbf{k}}}}\left(-\frac{vk^{*}}{m+E_{\mathbf{k}}},1\right)^{T}$. We have used the notation $k=k_{x}+ik_{y}$ and $k^{*}=k_{x}-ik_{y}$. Because we have two bands, the Berry curvature takes the simple form, for the positive band:

\begin{eqnarray}
\mathcal{B}_{i,+}=i\frac{\epsilon^{ijl}\langle +|\partial_{k_{j}}H|-\rangle\langle -|\partial_{k_{l}}H|+\rangle}{4E^{2}_{\mathbf{k}}}.\label{Omegaplus}
\end{eqnarray}

After pretty simple algebra, we get

\begin{equation}
\mathcal{B}_{z,+}=-\frac{v^2}{2}\frac{m}{\left(v^2|\mathbf{k}|^2+m^2\right)^{\frac{3}{2}}}\label{Omegaplus2}
\end{equation}

From (\ref{Omegaplus}) it is easy to find $\mathcal{B}_{z,+}=-\mathcal{B}_{z,-}$. We then end up with the expected result (\ref{CSQED3}) (as usual $\hbar=1$):

\begin{equation}
\sigma_{H}\equiv e^{2}\gamma_{-}=\frac{e^{2}v^2}{2}\int \frac{d^{2}\mathbf{k}}{4\pi^2}\frac{m}{\left(v^2|\mathbf{k}|^2+m^2\right)^{\frac{3}{2}}}=\frac{e^{2}}{4\pi}\frac{m}{|m|}.
\end{equation}

We have closed the circle and demonstrated that the common origin of the Hall conductance in the case of the QHE and $QED_{2+1}$ is the nontrivial Berry phase in the momentum space picked by the eigenstates of the system. To end this introduction, we can comment on the  properties of $\mathcal{B}_{z}(\mathbf{k})$ under the effect of discrete symmetries, like time reversal symmetry and parity. If the system is time reversal invariant, one can show that $\mathcal{B}_{z}(-\mathbf{k})=-\mathcal{B}_{z}(\mathbf{k})$ while if the system is parity invariant the Berry curvature satisfies $\mathcal{B}_{z}(-\mathbf{k})=\mathcal{B}_{z}(\mathbf{k})$. These two relations imply that if the system is time reversal invariant, the Berry connection does not need to be zero, but the integral of the Berry curvature, i. e., the Hall conductivity, vanishes. 
Also, if the system is both time reversal invariant \emph{and} parity invariant, $-\mathcal{B}_{z}(\mathbf{k})=\mathcal{B}_{z}(-\mathbf{k})=\mathcal{B}_{z}(\mathbf{k})$, so the Berry curvature is zero.

\subsubsection{Haldane model and the valley Hall effect.}

Now it is turn to look for situations different to the already analized QHE where the system displays similar physics. The condition of vanishing Hall conductivity for time reversal invariant systems might seem too strong and one might not expect the physics described by the Chern Simons term in any time invariant system. However, with the aid of graphene as an ideal laboratory to play with, we will see that this is actually not the case and we can still find time reversal invariant system displaying nontrivial topological features. For the details and a for an exhaustive list of references we refer to the excelent reviews on the subject\cite{XCN10,HK10,QZ11}.

Let us consider how we can modify the lattice hamiltonian describing the low energy spectrum in graphene. We know that, if we do not consider the real spin, the low energy hamiltonian of graphene is described by the hamiltonian (\ref{effDirac}). If we insist in opening gaps for the whole set of bands, we can add to (\ref{effDirac}) the following matrices, $M_{1}=1_{\tau}\sigma_{z}$, $M_{2}=\tau_{z}\sigma_{z}$, and the two matrices $M_{3,4}=\tau_{x,y}\sigma_{x}$. Physically, $M_{1}$ represents an atomic energy invalance between the atoms belonging to the two sublattices in the honeycomb lattice. In graphene it might appear due to a mismatch between the graphene sheet and the substrate lattice or in the boron nitride. $M_{3,4}$ physicaly correspond to two degenerate realizations of the coupling between the electrons and a Kekul\'{e} distortion (a kind of optical phonon in the honeycomb lattice). The Kekul\'{e} distortion is a quite interesting coupling by its own right. In its vector form $H_{Kekul\acute{e}}=\bm\Delta(\mathbf{r})\bm\tau\sigma_{x}$ it constitutes a vector coupling with the electronic structure at low energies. This coupling iis symmetric under chiral rotations of the form $U(\theta)=e^{i\tau_{z}\theta/2}$ if the $\bm\Delta(\mathbf{r})$ rotates as a vector in two dimensions. If the coupling $H_{Kekul\acute{e}}$ is generated by any form of interaction, the particular choice of $\bm\Delta(\mathbf{r})$ implies an spontaneous chiral symmetry breaking in graphene. As long as we are at finite temperature, increasing $T$ would eventually lead to a chiral symmetry restoring by the appearance of defects in the vector field $\bm\Delta(\mathbf{r})$, through a Kosterlitz-Thouless type of transition, which has some topological concepts behind, as we commented in the introduction. In terms of the Berry phase we are discussing here, this coupling is trivial.

The first, third and fourth matrices are time reversal invariant, as long as the time reversal operation acts in the following way on the tight binding hamiltonian: $\Theta H(\mathbf{K})\Theta^{-1}=H^{*}(-\mathbf{K})$, that is, it interchanges the two Dirac species and takes the complex conjugate. The only matrix that breaks time reversal symmetry is $M_{2}$. Haldane devised a tight binding hamiltonian which in the low energy sector a mass like $M_{2}$ appears, by adding complex second neighbour hoppings which opposite value in the two sublattices and an unbalance site energy between the two sublattices of the honeycomb lattice.

Let us make extensive use of the calculation performed in the previous section. The haldane model in the low energy sector reads:

\begin{equation}
H=v\tau_{z}\sigma_{x}k_{x}+v1_{\tau}\sigma_{y}k_{y}+m\tau_{z}\sigma_{z}.\label{Haldaneeff}
\end{equation}

Because there is no mixing between the two Dirac species, we can apply the formula (\ref{Omegaplus}) to each specie separately. For the first specie we find (\ref{Omegaplus2}). The important part comes now. The eigenstates for the two species are related by $|2,+\rangle =i\sigma_{y}|1,+\rangle$ and $|2,-\rangle =-i\sigma_{y}|1,-\rangle$ while the velocity operators satisfy $v^{1,2}_{j}=\sigma_{y}\nabla_{k_{j}}H^{2,1}(\mathbf{k})\sigma_{y}$, where the superscript labels the two species. Taking into account the relations between the eigenstates and the velocity operators, one finds $\mathcal{B}^{(1)}_{z,+}(\mathbf{k})=\mathcal{B}^{(2)}_{z,+}(\mathbf{k})$. It means that after integrating over the momentum domain, both species equally contribute to the Hall conductivity, and we get $\sigma_{H}=\frac{e^2}{2\pi}sign(m)$. This is consequent with the fact that the system is not time reversal invariant.

What if we consider now the same model, but with $M_{1}$ instead of $M_{2}$? This time the eigenstates for the two species satisfy $|2,+\rangle=\sigma_{x}|1,-\rangle$ and $|2,-\rangle=\sigma_{x}|1,+\rangle$, that is, this time the positive and negative solutions for one specie are related to the negative and positive solutions for the other specie, respectively. Now, in terms of $\sigma_{x}$, the velocity operators for the two species now satisfy $\sigma_{x}\nabla_{j}H^{(2)}\sigma_{x}=-\nabla_{j}H^{(1)}$, so, putting all together in the expression for the Berry curvature, we find $\mathcal{B}^{(2)}_{z,+}=-\mathcal{B}^{(1)}_{z,+}$, that is, now both species have opposite $\mathcal{B}_{z,+}$, so the total Berry curvature is zero. 

It is interesting to note that, despite of the total Berry curvature and $\sigma_{H}$ are zero, the contribution of each specie to $\mathcal{B}_{z,+}$ does not. So, the two species have a nontrivial topological structure separately represented by a nonzero Berry curvature, as long as there is no coupling between them. Then the question is if there is any physical observable depending on the separate Berry phase for each specie.  In order to ask this question, we will make use of the Chern-Simons effective action formalism.

Let us denote the Hall conductivity of each specie by $\sigma_{K}$ and $\sigma_{K'}$. We know that $\sigma_{K}+\sigma_{K'}=0$ so $\sigma_{K'}=-\sigma_{K}\equiv-\sigma$. Let us postulate the existence of two independent $U(1)$ gauge fields, one being the usual electromagnetic field $A_{i}$ and a second one $V_{i}$. The gauge field $A_{i}$ will be coupled to the standard fermionic current $J^{\mu}$, while the vector $V_{i}$ will couple to a quiral current $J^{\mu}_{\tau}$, that couples with opposite sign to each Dirac specie. The gauged Dirac lagrangean now reads:

\begin{equation}
\mathcal{L}=\bar{\psi}_{\mathbf{k}}\left(\gamma^{0}\omega -v \bm\gamma\mathbf{k}-m1\right)\psi_{\mathbf{k}}-eJ^{\mu}A_{\mu}-eJ^{\mu}_{\tau}V_{\mu},\label{gaugedDirac}
\end{equation} 

with the obvious definition of the $\gamma$-matrices and $\bar{\psi}=\psi^{+}\gamma^{0}\equiv\psi^{+}M_{1}$. We can play the game done previously in the context of the $QED_{(2+1)}$ and calculate the effective one-loop action for the two $U(1)$ gauge fields integrating out the fermion fields.. Among all the terms in the effective action, there are two nonvanishing terms that have the structure of (pseudo) Chern-Simons terms:

\begin{eqnarray}
\mathcal{S}_{pcs}=\frac{1}{4\pi}(\sigma_{K}-\sigma_{K'})\int d^{3}x\epsilon^{\mu\nu\rho}\left(V_{\mu}\partial_{\nu}A_{\rho}+
A_{\mu}\partial_{\nu}V_{\rho}\right)+eJ^{\mu}A_{\mu}+eJ^{\mu}_{\tau}V_{\mu},\label{PCS1}
\end{eqnarray}
 where the couplings with the fermion currents are added by hand. The average of the quiral current is the found by taking the functional derivative of (\ref{PCS1}) with respect the chiral field $V_{\mu}$:
 
\begin{equation}
\langle J^{\mu}_{\tau}\rangle\equiv\langle J^{\mu}_{K}-J^{\mu}_{K'}\rangle=\frac{\delta\mathcal{S}_{cs}}{\delta V_{\mu}}=\frac{e}{\pi}\sigma\epsilon^{\mu\nu\rho}\partial_{\nu}A_{\rho}.
\end{equation}

This expression means that there is a nonzero transverse current reacting to an external electromagnetic field, which is actually an inbalance between currents belonging to different Dirac species. In order to make this statement more precise, let us employ the Callan-Harvey mechanism discussed above. Let us take linear combinations of the fields $A_{\mu}$ and $V_{\mu}$: $a_{\mu}=A_{\mu}+V_{\mu}$ and $b_{\mu}=A_{\mu}-V_{\mu}$, so $A_{\mu}=\frac{1}{2}(a_{\mu}+b_{\mu})$ and $V_{\mu}=\frac{1}{2}(a_{\mu}-b_{\mu})$, and we do the same for the currents: $J^{\mu}=J^{\mu}_{K}+J^{\mu}_{K'}$ and $J^{\mu}_{\tau}=J^{\mu}_{K}-J^{\mu}_{K'}$. Writing (\ref{PCS1}) in terms of $a_{\mu}$ and $b_{\mu}$ we have:

\begin{equation}
\mathcal{S}_{pcs}=\frac{1}{4\pi}\sigma\int d^{3}x\epsilon^{\mu\nu\rho}\left(a_{\mu}\partial_{\nu}a_{\rho}-b_{\mu}\partial_{\nu}b_{\rho}\right)+eJ^{\mu}_{K}a_{\mu}+eJ^{\mu}_{K'}b_{\mu},\label{PCS2}
\end{equation}

That is, two independent Chern-Simons terms with opposite sign. We note that the corresponding Hall conductivity is the Berry phase for each single Dirac specie. Now put the system described by (\ref{PCS2}) in a finite region. Applying the Callan-Harvey mechanism, we find that the system develops two counterpropagating chiral currents belonging to different Dirac species. This is known in the context of graphene as the valley Hall effect.

In the case  of the QHE, we found that the system developed chiral one dimensional propagating states at the boundaries in a number given by the value of $\sigma_{H}$. Because the Hall conductivity can have any integer value, we can have the situation of getting any integer value of chiral states. In the case of $QED_{(2+1)}$ and the valley Hall effect, this is not possible, because $\sigma_{H}$ can only take two values $\sigma_{H}=\pm1$ in appropriate units. This is another important difference between the QHE and $QED_{(2+1)}$ and time reversal invariant topological insulators: the topological invariant in the two later cases is a 
$Z_2$ invariant. We note by passing that in the Haldane model, the topological invariant is also a $Z_2$ invariant, and we have one chiral state propagating along the same direction at the boundaries.

We can make a comment concerning the robustness of the edge states in these systems. In the case of the QHE, all the edge states present in the system are propagating in the same direction, independently of their number. It means that those states cannot scatter backwards, so they are dissipationless and the Hall conductivity remains quantized even in disordered realistic samples. The same happens in the Haldane model. In the case of the valley Hall effect, however, we have pairs of counterpropagating edge states, but belonging to different Dirac species, so in principle the edge states can backscatter if there is any perturbation or interaction coupling the two Dirac species, so the system is not protected against all kinds of perturbations, as it happens in the QHE. However, it turns out to be that the backscattering rate between species is extremely small as it can be seen in the case of impurities (cite paper of paco/gladys) and interactions (cite me/Henning). 
We then conclude that we can find out the way of unraveling the nontrivial topological structure of some time reversal invariants by using the Berry phase. Here we took the route of analysing the system in terms of an effective pseudo Chern Simons theory, but, as it happened with the QHE, we can study the topological structure of the system by means of the behavior of the Bloch wave functions in the Brillouin zone. This has been done in (cite all). We will not pursue this line here.

\subsubsection{Quantum spin Hall effect.}

In graphene, when spin is taken into account, the possibilities of mass-like couplings notably increase. Among all of these possible couplings, there is one that is of paramount importance, since is the on that will be generally responsible of the Quantum Hall effect in general time reversal invariant materials, different from graphene. The physical effect involving spin that will lead to the coupling we are talking about is the spin-orbit coupling. The spin-orbit coupling, as its name points out, couples the spin momentum with the orbital motion of the electrons in the lattice. It takes the general form of $H_{so}=\lambda\mathbf{S}\cdot\mathbf{L}$, with $\lambda$ some constant depending on the microscopic details of the atomic potential. It is easy to see that the term $H_{so}$ respects time reversal symmetry because both operators $\mathbf{S}$ and $\mathbf{L}$ flip their directions under reversing the time. The low energy effective spin-orbit coupling in graphene takes the form (we refer the literature for explicit lattice derivations: cite dresselhaus(1965) and others):

\begin{equation}
H_{so}=\Delta_{so}\sigma_{z}\tau_{z} s_{z}.\label{SOcoupling}
\end{equation}  

It is interesting to note that the matrix structure of (\ref{SOcoupling}) is similar to $M_{2}$ in the Haldane model. In the Haldane model, the mass does not respect time reversal symmetry, but, because the spin changes its sign under time reversal invariant, the product $M_{2}s_{z}=\sigma_{z}\tau_{z} s_{z}$ as a whole is time reversal invariant. It is then clear that the low energy hamiltonian for graphene in addition to the term (\ref{SOcoupling}) consist in two time reversal related copies of the Haldane model. From the discussion of the Haldane model, we infer that the term (\ref{SOcoupling}) opens a gap for the two spin components, and, because the effective field theory in the case of the Haldane model possesses a genuine $	\mathbf{Z}_{2}$ Chern Simons term, the low energy effective field theory in the case of graphene with an spin-orbit term will have a term similar to (\ref{PCS2}), but this time each current does not correspond to different Dirac species, but they belong to different spin projection. It means that in a confined geometry, the spin-orbit term will induce at each boundary, two counterpropagating edge states with opposite spin projection, so if one has a wide sample and is capable to manipulate the system at the scale of the edges, under the effect of an appropriate bias voltage, he or she will lead an spin filtered current, which constitutes the holy grial in the field of spintronics.

Despite of this so much appealing property, the spin-orbit constant $\Delta_{so}$ in graphene turns out to be extremely small. In units of temperature, $\Delta_{so}$ is estimated to be of the order of $0.01$ K, so any other effect in graphene is likely to have a bigger characteristic scale and be dominant compared with $H_{s-o}$. Also, in graphene there is other coupling that mixes spin and orbital movement. It is known as the extrinsic or Rashba spin-orbit coupling, which has the form $H_{R}=\lambda_{R}(\mathbf{s}\times\mathbf{p})\cdot \hat{\mathbf{e}}_{z}$. The Rashba coupling takes the following form in the low energy Hamiltonian of graphene:

\begin{equation}
H_{R}=\lambda_{R}\left(\tau_{z}\sigma_{x}s_{y}-1_{\tau}\sigma_{y}s_{x}\right).\label{Rashbacoupling}
\end{equation}

The Rashba term, depending on $s_{x}$ and $s_{y}$ competes with the spin-orbit term (\ref{SOcoupling}), and it can be found that the system remains to be an insulator if $\lambda_{R}<\Delta_{so}$, so while this condition is fulfilled, the spin Hall conductivity remains nonzero and the system is a topological insulator. We have to mention that in this case, although the spin Hall conductivity is nonzero, it does not generally take quantized values \cite{CGV10}:

\begin{equation}
\sigma_{sH}=\frac{e^2}{4\pi}\frac{\Delta_{so}}{\lambda_{R}}\ln\left| \frac{\lambda_{R}+\Delta_{so}}{\lambda_{R}-\Delta_{so}}\right|.\label{nonquantizedsigma}
\end{equation}

It is interesting to note that in order to induce backscattering between the two spin filtered edge states at the boundary, we must have a perturbation that couples the two spin polarizations, that is, a time reversal breaking perturbation. This is a manifestation of the fact that the edge states are protected against backscattering by time reversal invariance. 

The hamiltonian (\ref{SOcoupling}) together with the continuum effective Dirac hamiltonian for graphene are useful for describing the physics we are interested to in a simple way, but of course not all the two dimensional topological insulators must be described at the lattice level by this hamiltonian, although the effective field theory contains a Chern Simons term. We saw an example of this in the lattice description of the QHE. Indeed in the case of graphene, the discussion made so far remains valid as long as the value of the gap is not too big compared with the bandwidth, and the Berry curvature concentrates around the Dirac points. For these situations where the Dirac approximation is not the correct one, one may take to routes: to construct a fairly good effective hamiltonian for the electrons that captures the topological features of the system, or to develop a general theory for the $Z_{2}$ topological invariant in time reversal invariant systems taking into account all the information contained in the Brilloin zone.

The second route is being subject of current research and several methods have been proposed to construct such topological invariant. We will not review them here and we refer to the literature\cite{KM05,MB07,FH07,QHZ08,R09}. Concerning the first route, we will use bilayer graphene as an example. 

\subsubsection{Topological insulating phases in bilayer graphene.}

For bilayer graphene in the Bernal stacking the unit cell is made of two pairs of carbon atoms, each pair belonging to a different graphene layer. The simplest lattice hamiltonian that shows nontrivial topological features is the one where only hopping amplitudes are allowed between the most adjacent atoms in the two layers. If we define the wavefunction of the system for the low energy effective model around the $K$ point by $\Psi_{\mathbf{k}}=\left(a_{1,\mathbf{k}},b_{1,\mathbf{k}},a_{2,\mathbf{k}},b_{2,\mathbf{k}}\right)^{T}$ being the components the amplitudes in the sublattices $a$ and $b$ of the two layers, this simplest model reads $H=\sum_{\mathbf{k}}\Psi^{+}_{\mathbf{k}}\mathcal{H}_{\mathbf{k}}\Psi_{\mathbf{k}}$ with
\begin{equation}
\mathcal{H}_{\mathbf{k}}=\left(\begin{array}{cc}
\bm\sigma\mathbf{k} & t_{\perp}\sigma_{-} \\
t_{\perp}\sigma_{+} & \bm\sigma\mathbf{k}  
\end{array}\right)	,\label{BiHam}
\end{equation} 
where we have used the standard definition, $\sigma_{\pm}=\left(\sigma_{x}\pm i\sigma_{y}\right)/2$, and $t_{\perp}$ is the hopping amplitude between the sublattices $a_{2}$ and $b_{1}$.  The effective model for the other $K$ point is found from (\ref{BiHam}) by substituting $k_{x}$ by $-k_{x}$. In what follows we will use the following convention for the different Pauli matrices appearing in the model. The $\bm\sigma$ matrices stand for sublattice, $\bm\tau$ stand for $K,K'$ points, $\bm\mu$ stand for layer index, and $\mathbf{s}$ refers to the spin quantum number. The system consists then in four bands around each $K$ point and are degenerate in spin, i.e., sixteen bands for the whole effective low energy model.

In absence of any other coupling or perturbation and for energies much smaller than $t_{\perp}$ the model can be simplified further to an effective two band model for each $K$ point. As described in Sect.~\ref{sec_bil}, this effective model for the amplitudes $\left(a_{2,\mathbf{k}},b_{1,\mathbf{k}}\right)^{T}$ is
\begin{equation}
\mathcal{H}^{eff}_{\mathbf{k}}=\frac{v^2}{t_{\perp}}\bm\sigma\mathbf{k}\sigma_{x}\bm\sigma\mathbf{k}=\frac{v^2}{t_{\perp}}\left(\begin{array}{cc}
0 & (k_{x}-ik_{y})^{2}\\
(k_{x}+ik_{y})^{2} & 0
\end{array}\right).\label{BeffHam}
\end{equation}

In some cases the model (\ref{BeffHam}) is enough to analyze the topological structure of the system. This is the case, for instance, when an external homogeneous magnetic field is applied. The low energy part of the Landau spectrum of the system can be derived directly from (\ref{BeffHam}) to get $E_{n}=\pm\omega_{l}\sqrt{n(n-1)}$ \cite{MF06}, that is, each valley and spin projection contributes with two zero modes instead of one, as it happened for monolayer graphene. This change can be experimentally seen in the value of the Hall conductivity $\sigma_{H}$ for bilayer graphene at zero carrier concentration, being the value of $\sigma_{H}$ twice the value in monolayer graphene \cite{NMetal06}.

Another interesting effect is the application of electrostatic potential difference between the two layers. This effect is modeled by the term $V=V_{0}1_{s}\mu_{z}1_{\tau}1_{\sigma}$ in (\ref{BiHam}) or $V=V_{0}\sigma_{z}$ in the effective model (\ref{BeffHam})\cite{M06}. By making use of (\ref{Berryconnection}) and the eigenstates of $\mathcal{H}_{\mathbf{k}}+V$ it can be shown that the Hall conductivity of bilayer graphene in presence of a voltage difference is zero as long as this voltage difference respects the time reversal inversion symmetry, but as it happened with monolayer graphene, we can still have a valley Hall effect with value of the valley Hall conductivity to be $\sigma_{K}=\frac{2e^2}{2\pi}sign(V_{0})$. In this case, it means that after applying a voltage difference between the layers a pair of counterpropagating edge states polarized in valley appear at the edge of the sample. This is quite appealing because is by far easier for experimentalists to apply an external voltage difference between the layers in bilayer graphene than opening a gap of the type $M_{1}$ in monolayer graphene. Concerning spin, we expect to find similar topological structure than in monolayer graphene when the intrinsic and Rashba spin orbit terms are considered now. The most general term involving both effects in bilayer graphene can be found to be\cite{G10,MK10}:

\begin{equation}
H_{so}=\lambda_{1}1_{\mu}\tau_{z}s_{z}\sigma_{z}+\lambda_{2}\mu_{z}\tau_{z}s_{z}1_{\sigma}+\lambda_{3}\left(\mu_{z}1_{\tau}s_{x}\sigma_{y}-\mu_{z}\tau_{z}s_{y}\sigma_{x}\right)+
\lambda_{4}\left(\mu_{y}1_{\tau}s_{x}\sigma_{z}-\mu_{x}\tau_{z}s_{y}\sigma_{x}\right).\label{BiSo}
\end{equation} 

In the hamiltonian (\ref{BiSo}) the term with $\lambda_{1}$ is the equivalent version of the spin-orbit term in monolayer graphene, while the term with $\lambda_{2}$ is proper from bilayer graphene. Both terms open nontrivial gaps in the bilayer spectrum and lead to a quantized value for the spin Hall coefficient, $\sigma_{s}=\frac{2e^{2}}{2\pi}sign(\lambda_{1}+\lambda_{2})$. The terms associated to $\lambda_{3}$ and $\lambda_{4}$ are Rashba terms which destroy the quantization of the spin Hall response in a similar manner that happens in monolayer graphene.

An interesting situation appears when the intrinsic spin-orbit terms and the V term are considered simultaneously. The system displays both Spin and Valley Hall effects, being both Hall conductances quantized, $$\sigma_{s}=\frac{e^{2}}{2\pi}\left(sign(\lambda_{1}+\lambda_{2}+V_{0})+sign(\lambda_{1}+\lambda_{2}-V_{0})\right),$$ and $$\sigma_{K}=\frac{e^{2}}{2\pi}\left(sign(\lambda_{1}+\lambda_{2}+V_{0})+sign(V_{0}-\lambda_{1}-\lambda_{2})\right).$$ As usually happens, the quantized values for the distinct Hall responses are associated to the fact that a gap is opened in the system. Of course bilayer graphene is not an exception and the presence of a finite Fermi surface makes any Hall response to be nonquantized.

To finish this section, we can comment on the fate of the edge states when gated graphene bilayer is considered. As we have said the system shows a nonzero valley Hall conductivity $\sigma_{K}$ so if we apply the Callan-Harvey mechanism to this situation we will find two conuterpropagating edge states (without taking into account the real spin) filtered in the valley number\cite{CPS07}. As it happened in monolayer graphene or in the QHE, the topological number gives us only the number of propagating quiral states, but not their velocities, being them dependent on the microscopic details. It might well happen that both counterpropagating chiral states posses different velocities. This fact have important consequences for the transport properties, because the backscattering rates when disorder or interactions are considered strongly depends on these values\cite{COS10}. 

\section{Summary and future directions.}
One of the purposes of this review, besides presenting some QFT aspects of condensed matter in a language available to high energy colleagues, was to explore the theoretical aspects of the problem  that can still be developed in the near future. The giant explosion of graphene publications after its synthesis has explored all possible aspects of it and the latest activity is mostly centered on the  applications, devices, and exploring standard condensed matter phenomena  that are no very different in graphene than in a usual electron gas.

It is fair to say that most of the special properties of the material originate at the Fermi points and as such are difficult to access. New suspended samples allow for the measurement of the electronic properties within a range of $\rho \approx 10^8$cm$^{-2}$ or $\epsilon_F \approx 1$meV in single layer graphene. The density dependence of the effective mass and Fermi velocity at low carrier concentrations showed an enhancement of the Fermi velocity of almost a factor of three. This effect is consistent with the expected renormalization of the Fermi velocity in the absence of screening, given a graphene's "fine structure constant" $\alpha = e^2 / v_F \approx 2.3 - 2.5$. Similar estimates for $\alpha$ have been used  to explain the observation of plasmaron satellites in samples with high carrier concentrations. This value of $\alpha$ puts single layer graphene into the intermediate to strong coupling regime, although its flow towards zero takes the system to the weak coupling regime at low energies.

The issue of opening a gap in monolayer graphene is one of the few fundamental questions that still remain controversial in the topic.  Some recent lattice gauge theory calculations indicate that the physical parameters of graphene can be at the edge of the symmetry broken region. Given the interest of the subject on the theoretical side and even more for the practical applications, a special effort has to be devoted to develop more sophisticated calculations in graphene. We must although warn that there are at the present no experimental evidences favoring  a gap opening from interactions in the intrinsic monolayer system that has been tested at incredibly low energies \cite{EGetal11}.

There are some QFT aspects of graphene of recent interest that have been left aside of this review not because they are not interesting and worth to pursue but for lack of spacetime, or of expertise, or because we consider that the field is not yet mature.
A very interesting aspect that is being object of great attention and will probably be developed in the near future is that of the the ADS/CFT correspondence \cite{Sach10} which has been studied in the context of graphene in \cite{MSF09}. The search for charge fractionalization both for purely theoretical reasons as well as for potential application in the field of quantum computing has also used graphene as a simple example to show the main features \cite{J07}. The same can be said on the search for Majorana fermions, another exploding subject in condensed matter \cite{W09} with a clear relation with its high energy counterpart. 

\section{Acknowledgments} Support from MEC (Spain) through grants FIS2008-00124, PIB2010BZ-00512  is acknowledged.

\section*{References}
\bibliography{JPA}

\begin{thebibliography}{180}
\bibitem[{FQH()}]{FQHE}
\bibinfo{note}{Note that this is a phenomenon where the electron kinetic is
  completely quenched, and the only energy scale is $e^2 / ( \epsilon \ell_B
  )$, where $e$ is the electric charge, $\epsilon$ is the dielectric constant,
  and $\ell_B$ is the magnetic length.}
\end{thebibliography}

\end{document}